\documentclass[a4paper,fleqn,usenatbib]{mnras}

\usepackage{newtxtext,newtxmath}

\usepackage[T1]{fontenc}
\usepackage{ae,aecompl}
\usepackage{subfig}

\usepackage{graphicx}	
\usepackage{amsmath}	
\usepackage{amssymb}	

\usepackage{color}
\usepackage{BibDef}
\def\lsim{\mathrel{\rlap{\lower 3pt \hbox{$\sim$}} \raise 2.0pt \hbox{$<$}}}
\def\gsim{\mathrel{\rlap{\lower 3pt \hbox{$\sim$}} \raise 2.0pt \hbox{$>$}}}
\def\msun{\rm {M_\odot}}

\def\mach{\mathcal{M}}

\def\vecdist{\mathbf{x}_{j,{\rm s}}}
\def\nhtot{n_{\rm H_{tot}}}
\def\grav{\rm G}

\title[H$_2$ chemistry with mechanical supernova feedback] 
{H$_2$ chemistry in galaxy simulations: an improved supernova feedback model}
\author[A. Lupi]{Alessandro Lupi$^{1}$\thanks{E-mail:
lupi@iap.fr}\\
$^1$Sorbonne Universit\`{e}s, UPMC Univ Paris 6 et CNRS, UMR 7095, Institut d'Astrophysique de Paris,\\ 98 bis bd Arago, F-75014 Paris, France
}
\begin{document}

\date{Draft \today}

\pagerange{\pageref{firstpage}--\pageref{lastpage}} \pubyear{2018}

\maketitle

\label{firstpage}

\begin{abstract}
In this study, we present and validate a variation of recently-developed physically motivated sub-grid prescriptions for supernova feedback that account for the unresolved energy-conserving phase of the bubble expansion. Our model builds upon the implementation publicly available in the mesh-less hydrodynamic code \textsc{gizmo}, and is coupled with the chemistry library \textsc{krome}. Here, we test it against different setups, to address how it affects the formation/dissociation of molecular hydrogen (H$_2$). First, we explore very idealised conditions, to show that it can accurately reproduce the terminal momentum of the blast-wave independent of resolution. Then, we apply it to a suite of numerical simulations of an isolated Milky Way-like galaxy and compare it with a similar run employing the delayed-cooling sub-grid prescription. We find that the delayed-cooling model, by pressurising ad-hoc the gas, is more effective in suppressing star formation. However, to get this effect, it must maintain the gas warm/hot at densities where it is expected to cool efficiently, artificially changing the thermo-chemical state of the gas, and reducing the H$_2$ abundance even in dense gas. Mechanical feedback, on the other hand, is able to reproduce the H$_2$ column densities without altering the gas thermodynamics, and, at the same time, drives more powerful outflows. However, being less effective in suppressing star formation, it over-predicts the Kennicutt-Schmidt relation by a factor of about 2.5. Finally, we show that the model is consistent at different resolution levels, with only mild differences. 

\end{abstract}

\section{Introduction}
According to the current cosmological model, baryons cool down and fall within the potential well of dark matter haloes, fragmenting and forming stars (and galaxies). From simple arguments, star formation (SF) should occur on a free-fall time--scale, consuming very quickly all the available gas supply \citep[e.g.,][]{bournaud10,dobbs11}. However, the typical observed time--scale for SF is much longer than that expected from these simple arguments. One of the possible reasons behind this difference is stellar feedback, which evacuates the gas from the SF sites, suppressing the actual SF efficiency. 
One of the most important feedback processes to be considered is supernova (SN) feedback, i.e. the explosion of massive stars ($M_{\rm s} > 8\,\msun$) and accreting white dwarfs in binary systems (as type Ia SNe).

Although an accurate physical description of the SN explosion mechanism is still missing, at pc scales SN events can be simply modelled as an instantaneous injection into the surrounding medium of mass, metals, and energy. In the last few decades, many authors investigated the evolution of the SN-driven bubble, with both analytical calculations and numerical simulations \citep[e.g.,][]{chevalier74,mckee77,cioffi88,kimostriker15,martizzi15,geen16}. However, to properly capture this evolution in numerical simulations, in particular the initial energy conserving phase \citep[the Sedov--Taylor phase;][]{taylor50,taylor50b,sedov59}, when radiative losses are still unimportant, extreme mass and spatial resolution are needed. Unfortunately, this is currently not achievable in galaxy-scale simulations, and it gets even worse in cosmological ones.
In this case, a simple thermal energy injection would result in efficient radiative losses and negligible feedback effect, hence ad-hoc sub-grid prescriptions are necessary to overcome this problem.

At very low resolution, when the inter-stellar medium (ISM) is unresolved, empirical models are usually employed, like in (i) \citet{navarro93}, where energy is injected in the kinetic form, (ii) \citet{springel03,keller14}, where  the gas is described via a sub-grid multi-phase model, (iii) \citet{stinson06}, where cooling is shut off for some time to allow the blast-wave to expand following the energy-conserving solution, and (iv) \citet{dallavecchia12}, where many SN events are stochastically grouped together to produce more energetic explosions. 

Despite the success of these models, at higher resolution, when the multi-phase structure of the ISM can be resolved, a more physically motivated model is desirable. Recently, several authors put an effort to implement a new sub-grid supernova feedback model, so-called mechanical feedback, in different hydrodynamic codes \citep{hopkins13sf,kimm14,smith18,hopkins18}, where the feedback deposition takes into account the possibly unresolved Sedov--Taylor phase, injecting 1) the initial ejecta momentum/thermal energy if this phase is resolved, or 2) the terminal momentum of the bubble otherwise.

Recently, \citet{martizzi15} investigated the evolution of SN bubbles in homogeneous and inhomogeneous media with very high resolution, in order to properly model the Sedov--Taylor phase. With their simulations, the authors were able to fit an analytic formula to properly describe the thermal energy and the momentum of the SN bubble at different stages of the evolution. This prescription, so far implemented only by \citet{semenov17}, allows a more accurate estimation of the amount of thermal energy and momentum to be injected as a function of the `coupling radius', i.e. the resolution of the simulation. Here, we implement the same prescription in the numerical code \textsc{gizmo} \citep{hopkins15}, as a variation of the already implemented mechanical feedback model by \citet[][H18 hereafter]{hopkins18b}, and we validate it against different tests, i.e. a single explosion in a uniform medium, and an isolated galaxy. 

Here, we follow the evolution of a single galaxy with a complete non-equilibrium reduced network of H$_2$ and a physically motivated SN feedback, similarly to what is done in \citet{hu18} in the case of a dwarf galaxy, \footnote{Although \citet{kimm17} performed radiation-hydrodynamics cosmological simulations including non-equilibrium H$_2$ chemistry, their model does not have a full out-of-equilibrium treatment, since they do not directly follow H$^-$ and H$_2^+$. In particular, they assume that H$^-$ is in collisional equilibrium and that H$_2^+$ is instantaneously dissociated, and this is different from what we employ here.} to assess whether the correlation between H$_2$ and star formation rate (SFR) surface densities can still be naturally reproduced without assuming a priori dependence of SF on the H$_2$ abundance \citep{lupi18a}.

The manuscript is organised as follows: in Section~\ref{sec:model}, we describe the model and its implementation in the code; in Section~\ref{sec:test} we validate it by means of a single SN explosion test in a pseudo-uniform medium; in Section~\ref{sec:setup}, we describe the sub-grid model employed in the galaxy simulations; in Section~\ref{sec:results}, we present our results; in Section~\ref{sec:conclusions}, we discuss the limits of the model and we draw our conclusions.

\section{Mechanical feedback model}
\label{sec:model}

\subsection{The hydrodynamic code \textsc{gizmo}}
\textsc{gizmo} \citep{hopkins15}, descendant from \textsc{Gadget3}, itself descendant from \textsc{Gadget2} \citep{springel05}, implements a new method to solve hydrodynamic equations, aimed at capturing the advantages of both usual techniques used so far, i.e. the Lagrangian nature of smoothed particle hydrodynamics codes, and the excellent shock-capturing properties of Eulerian mesh-based codes, and at avoiding their intrinsic limits.
The code uses a volume partition scheme to sample the volume, which is discretised amongst a set of tracer `particles' which correspond to unstructured cells. Unlike moving-mesh codes \citep[e.g. \textsc{arepo};][]{springel10}, the effective volume associated to each cell is not defined via a Voronoi tessellation, but is computed in a kernel-weighted fashion. Hydrodynamic equations are then solved across the `effective' faces amongst the cells using a Godunov-like method, like in standard mesh-based codes. In these simulations, we employ the mesh-less finite-mass method, where particle masses are preserved, and the standard cubic-spline kernel, setting the desired number of neighbours to 32.
Gravity is based on a Barnes-Hut tree, as in \textsc{Gadget3} and \textsc{Gadget2}. It provides two different softening schemes, fixed and adaptive, for all particles. While a fixed gravitational softening can be a reasonable assumption for dark matter and stars, gas is volume-filling, hence adaptive softenings represent a better choice. Moreover, adaptive softenings ensure that both hydrodynamics and gravity are computed assuming the same mass distribution within the kernel, ensuring that the force is always Newtonian.

\subsection{Neighbour search and coupling}
Here we report the basic assumptions of the mechanical feedback model publicly available in \textsc{gizmo}, presented in H18, for sake of completeness, and we highlight the differences with our new prescription. 
 
When a SN explodes, we must couple energy (and/or momentum) and mass (and metals) to the surrounding medium in a conservative way. Unlike in mesh-based codes, where the volume is perfectly sampled, and the coupling is almost straightforward, the coupling in particle-based codes like \textsc{gizmo} requires more care. The algorithm by H18 considers an effective number of neighbours around every star particle $N_\star = 4\upi/3 h_{\rm s}^3 n_{\rm s}^3$, where $n_{\rm s}=\sum_jW(\vecdist, h_{\rm s})$, $W$ is the kernel function, $\vecdist=\mathbf{x}_j-\mathbf{x}_{\rm s}$ is the particle separation vector, and $h_{\rm s}$ is the star kernel size. In order to guarantee that the area around the star is covered, $h_{\rm s}$ {is  determined} by requiring $N_\star=64$.

To better couple the feedback with the gas, instead of distributing it to the gas within the kernel of the star only, H18 also considers all gas particles with their kernel encompassing the star, i.e. $|\vecdist| < h_j$, with $j$ representing the $j$-th particle.

\begin{figure}
\includegraphics[width=\columnwidth]{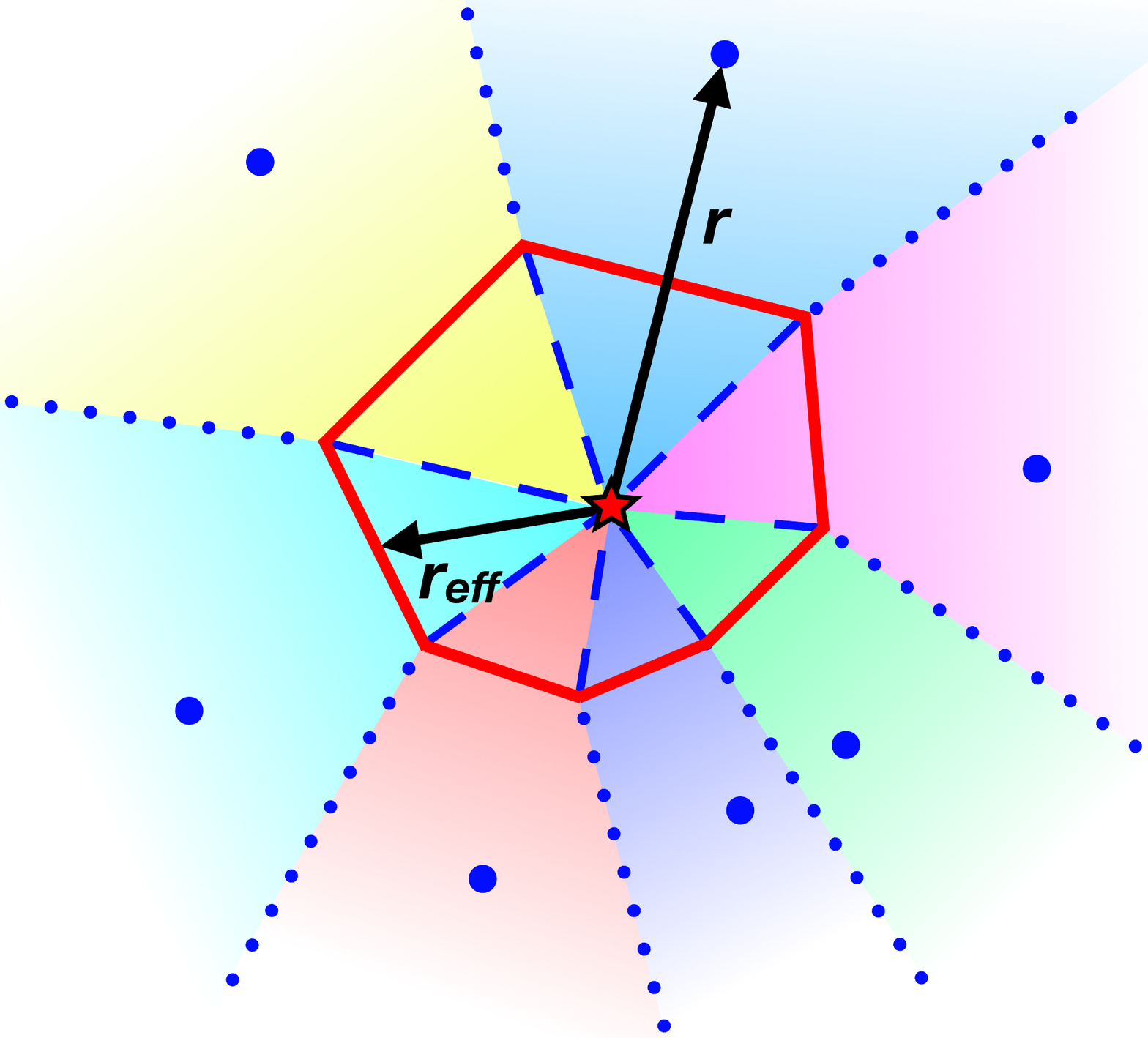}
\caption{Coupling scheme for the mechanical SN feedback, adapted from H18, where the original version of the algorithm is presented. The red lines correspond to a convex hull around the stellar particle, with each line representing an effective face shared with a gas cell.  The dotted (dashed) lines correspond to the separation between the volumes associated to each gas neighbour, outside (inside) the convex hull built around the star. All the separation lines between particle pairs (including the stars) are obtained here from a Voronoi tessellation, that is similar to the result obtained in \textsc{gizmo}. Each neighbour receives an amount of momentum proportional to the renormalised vector weight, computed from the solid angle subtended by the face. In our variation of the original model by H18, the arrow identified with $r$ here represents the coupling radius used to compute the bubble expansion solution, whereas that identified with $r_{\rm eff}$ corresponds to the effective distance used to prevent unphysical kicks to particles whose face is outside the maximum extension radius of the bubble for very-low density medium.}
\label{fig:scheme}
\end{figure}

After having identified the neighbours, the fraction $\omega_j$ of energy/momentum and mass each neighbour should receive is estimated as the normalised solid angle subtended by the gas element around the star. This is done by building a set of effective faces shared between star and gas with some convex hull, as shown in Fig.~\ref{fig:scheme}. Analytically, one can write
\begin{equation}
\omega_j = \frac{\Delta \Omega_j}{4\upi} = \frac{1}{2}\left(1-\frac{1}{\sqrt{1+\mathbf{A}_j\cdot {\vecdist}/(\upi |\vecdist|^3)}}\right),
\end{equation}
where $\mathbf{A}_j$ is the effective face area, which goes from $\omega_j\approx 1/2$ for $|\vecdist|^2\ll A_j$ to $\omega_j \approx A_j/(4\upi |\vecdist|^2)$ for $|\vecdist|^2\gg A_j$. Although different ways of constructing a convex hull exist, H18 assumes 

\begin{equation}
\mathbf{A}_j = \left\{n_j^{-2}\frac{\partial W(\vecdist,h_j)}{\partial |\vecdist|} + n_{\rm s}^{-2}\frac{\partial W(\vecdist,h_{\rm s})}{\partial |\vecdist|} \right\}\frac{\vecdist}{|\vecdist|}.
\end{equation}

We note that, when either $s$ or $j$ are outside the kernel of the other particle (W=0), this definition results in a constant area $\mathbf{A}_j$ that only depends on the properties of the particle with the largest kernel. For instance, in the extreme case of low density around the star, with all the gas neighbours having $h_j<|\vecdist|$ and $h_i\gg h_j$, this equation assigns equal weights to all the gas neighbours of the star, without taking into account their actual volume. \footnote{In some of our tests, this resulted in the spurious ejection of dense clumps around the star.} In these cases, unlike in H18, we employ an alternative definition of the effective area, 
\begin{equation}
\mathbf{A}_j = \frac{1}{n_i\,n_j}\left\{\frac{\partial W(\vecdist,h_j)}{\partial |\vecdist|} + \frac{\partial W(\vecdist,h_{\rm s})}{\partial |\vecdist|} \right\}\cdot \frac{\vecdist}{|\vecdist|}.
\end{equation}

Since momentum is also injected in the surroundings, one must guarantee that linear momentum is conserved. To do so, H18 imposes a {\it tensor} renormalisation, where the six-dimensional weights $\hat{\mathbf x}^\pm_{j,{\rm s}}$ are defined as
\begin{eqnarray}
\hat{\mathbf x}_{j,{\rm s}} &=& \frac{\vecdist}{|\vecdist|} = \sum_{+,-} \hat{\mathbf x}_{j,{\rm s}}^\pm\\
(\hat{\mathbf x}^+_{j,{\rm s}})^k &=& |\vecdist|^{-1} \max(\vecdist^k,0)\biggr|_{k=x,y,z}\\
(\hat{\mathbf x}^-_{j,{\rm s}})^k &=& |\vecdist|^{-1} \min(\vecdist^k,0)\biggr|_{k=x,y,z}.
\end{eqnarray}

The actual weight can therefore be derived as
\begin{eqnarray}
\mathbf{\tilde{w}}_j &=& \frac{\mathbf{w}_j}{\sum_i |\mathbf{w}_i|}\\
\mathbf{w}_j &=& \omega_j \sum_{+,-} \sum_k (\hat{\mathbf{x}}_{j,{\rm s}}^\pm)^k f_\pm^k\\
(f_\pm)^k &=& \left\{\frac{1}{2} \left[1+\left(\frac{\sum_i \omega_i |\mathbf{\hat{x}}_{i,{\rm s}}^\mp|^k}{\sum_i \omega_i |\mathbf{\hat{x}}_{i,{\rm s}}^\pm|^k}\right)^2  \right] \right\}^{1/2}.
\end{eqnarray}

With this definition, the right distribution of energy/momentum, mass, and metals is guaranteed, and, at the same time, the total linear momentum vector is ensured to vanish in the reference system of the star.

The fractional quantities each neighbour receives can now be defined as $\Delta m_j = |\tilde{\mathbf{w}}_j| m_{\rm ej}$, $\Delta m_{{\rm Z},j} = |\tilde{\mathbf{w}}_j| m_{\rm Z,ej}$, $\Delta E^*_j= |\tilde{\mathbf{w}}_j| E_{\rm SN}$, and $\Delta \mathbf{p}^*_j = \tilde{\mathbf{w}}_j p_{\rm SN}$. Here, $m_{\rm ej}$ is the ejecta mass, $m_{\rm Z,ej}$ is the metal mass in the ejects, $E_{\rm SN}$ is the energy injected by the SN, and $p_{\rm SN}$ the injected momentum. 

So far, all the used quantities were defined in the reference frame of the star ($S^*$). However, in a real galaxy, both stars and gas are in motion. In order to account for the gas-star relative motion and conserve both energy and momentum, we first move to the reference frame moving with the gas cell $S'$, and compute the pre-shock properties of the ejecta. The coupled mass and the metal mass are not affected by the boost, whereas the new ejecta momentum and energy become $\Delta \mathbf{p}_j' = \Delta \mathbf{p}^*_j + \Delta m_j (\mathbf{v}_{\rm s}-\mathbf{v}_j)$, and $\Delta E_j' = \Delta E^*_j + \frac{1}{2\Delta m_j}(|\Delta \mathbf{p}_j'|^2 - |\Delta \mathbf{p}^*_j|^2)$, respectively, where $\mathbf{v}_j$ is the gas velocity in the lab/simulation frame $S$. 
Now, we can update the $j$-th neighbour properties in $S'$ as
\begin{eqnarray}
\mathbf{p'}^{\rm new}_j =& \Delta \mathbf{p}_j'\\
E'^{\rm new}_j =& E_j' + \Delta E_j'.
\end{eqnarray}

Finally, we move back to the lab/simulation reference frame, where the new properties are written as

\begin{eqnarray}
m^{\rm new}_j &=& m_j + \Delta m_j\\
m^{\rm new}_{{\rm Z},j} &=& Z^{\rm new}_j m^{\rm new}_j = Z_j m_j + \Delta m_{{\rm Z},j}\\
\label{eqn:dp}\mathbf{p}^{\rm new}_j &=& m^{\rm new}_j\mathbf{v}_j + \Delta \mathbf{p}^\prime_j = \mathbf{p}_j + \Delta\mathbf{p}^*_j+\Delta m_j\mathbf{v}_{\rm s}\\
E^{\rm new}_j &=& E'_j + \Delta E'_j + \frac{1}{2m^{\rm new}_j}(|\mathbf{p}^{\rm new}_j|^2-|\mathbf{p}^{\prime\,\rm new}_j|^2) 
\end{eqnarray}
In $S'$, the pre-coupling total energy simply corresponds to $E'_j = U_j$, where $U_j$ is the gas cell internal energy. The updated internal energy of the gas cell in $S$ directly follows from the after-coupling total energy and momentum as
\begin{equation}
U^{\rm new}_j =  E^{\rm new}_j - \frac{1}{2m^{\rm new}_j}|\mathbf{p}^{\rm new}_j|^2 = U_j +\Delta E'_j - \frac{1}{2 m^{\rm new}_j}|\Delta \mathbf{p}^{\prime\,\rm new}_j|^2
\end{equation} 
where $\Delta E'_j - |\Delta \mathbf{p}_j^{\prime\,\rm new}|^2/(2 m^{\rm new}_j) \equiv \Delta U_j$.

\subsection{The unresolved energy-conserving phase}
\label{sec:momcons}
When resolution is low, the energy-conserving phase of the SN bubble expansion cannot be resolved. This inevitably leads to a quick loss of thermal energy due to radiative cooling, and a reduced effect of SN feedback. To avoid this, we account for the momentum gained during the unresolved phase as resulting from high resolution simulations of single SNe. In our model, we replace the terminal momentum and the thermal energy decay used by H18 following \citet{martizzi15}. In particular, we rewrite the thermal energy and the momentum in terms of the swept mass $m_{\rm swept}$ as
\begin{equation}
\begin{array}{lcl}
E_{\rm th}(m_{\rm swept}) &=&
\begin{cases} 
      E_{\rm th,ej} & m_{\rm swept}\leq m_{\rm cool} \\
      E_{\rm th,ej} \left(\frac{\min\{m_{\rm swept},m_{\rm r}\}}{m_{\rm cool}}\right)^{-\alpha/3}& m_{\rm swept} > m_{\rm cool} \\ 
\end{cases}\\
p(m_{\rm swept}) &=& p_0 \left(\frac{\min\{m_{\rm swept},m_{\rm cool}\}}{m_{\rm ej}}\right)^{0.5},
\end{array}
\label{eqn:ethpr}
\end{equation}
where, assuming an homogeneous medium with initial density $\rho_0$, $m_{\rm cool} = 4/3\upi \rho_0 r_{\rm cool}^3$, with $r_{\rm cool}$ the cooling radius, $\alpha$ is the declining slope for the thermal energy outside $r_{\rm cool}$, $m_{\rm r}=4/3\upi \rho_0 r_{\rm r}^3$ is the mass corresponding to the radius $r_{\rm r}$ after which the thermal energy is roughly constant, $E_{\rm th,ej}$ is the thermal energy during the Sedov--Taylor phase (corresponding to $\sim~0.69~E_{\rm ej}$), and $p_0=\sqrt{2m_{\rm ej}E_{\rm ej}}$ is the ejecta momentum, with $E_{\rm ej} \equiv \Delta E'$.

We notice that the equations in \citet{martizzi15}, those employed here are slightly different. First, instead of $m_{\rm ej}$, there is a scale radius $r_0$, used to extrapolate the momentum at small radii, corresponding to a scale mass $m_0=4/3\upi \rho_0 r_0^3$. However, the use of $m_0$ introduces an additional dependence of the terminal momentum on the initial ejecta momentum, which is not observed in similar studies of single SN explosions. Our choice removes this dependence. Second, in \citet{martizzi15}, the terminal momentum is reached at a radius $R_{\rm b}> R_{\rm cool}$. We performed several experiments and we found that, in our case, $R_{\rm cool}$ was a better choice (see Section~\ref{sec:test}) to reproduce the high-resolution simulations. 
Compared to H18, where the total mass of the $j$-particle is swept up, resulting in an instantaneous switch from the energy-conserving to the momentum-conserving solution, our model properly takes into account the continuous increase of the swept-up mass, with $m_j$ only representing an upper limit. As we will show in Section~\ref{sec:test}, this choice has an important impact on the terminal momentum, especially at high resolution.

According to \citet{haid16}, in the case of an inhomogeneous medium, the differences from the homogeneous case can be captured by considering each gas element separately, and determining the solution of the homogeneous case using the gas element properties. Following this approach, we can determine the right solution independently for each `cone' (see Fig.~\ref{fig:scheme}) by using the $j$-particle density and metallicity to estimate $\alpha$, $R_{\rm cool}$, and $R_{\rm r}$ as  \citep{martizzi15}
\begin{equation}
\begin{aligned}
&\alpha &=& 7.8 &\left(\frac{Z}{Z_\odot}\right)^{+0.050} &\left(\frac{\nhtot}{100\rm\, cm^{-3}}\right)^{+0.030},\\
&r_{\rm cool} &=& 3.0 &\left(\frac{Z}{Z_\odot}\right)^{-0.084} &\left(\frac{\nhtot}{100\rm\, cm^{-3}}\right)^{-0.42}\rm\, pc, \\
&r_{\rm r} &=& 5.5 &\left(\frac{Z}{Z_\odot}\right)^{-0.074} &\left(\frac{\nhtot}{100\rm\, cm^{-3}}\right)^{-0.40}\rm\, pc,\\
\end{aligned}
\end{equation}
where $\nhtot$ is the total number density of hydrogen nuclei and $\rho_0=\nhtot/X*m_{\rm H}$, with $X=0.76$ the Hydrogen nuclei fraction and $m_{\rm H}$ the Hydrogen mass. 
Using Eq.~\eqref{eqn:ethpr} and assuming $m_{\rm swept} = m_{\rm ej} + \min\{4/3\upi\rho_j|\vecdist|^3,m_j/|\tilde{\mathbf{w}}_j|\}$, we can estimate $\Delta U_j = E_{\rm th}(m_{\rm swept})$ and $\Delta \mathbf{p}''_j =p(m_{\rm swept})/p_0 \Delta \mathbf{p}'_j$, where $\Delta \mathbf{p}''_j$ replaces $\Delta \mathbf{p}'_j$ in Eq.~\eqref{eqn:dp}.

Although the model well captures the unresolved Sedov--Taylor phase and can reproduce the terminal momentum, as we will show in the next section, there are some cases where some neighbours can be  far away from the star, well beyond the radius at which the bubble should merge with the ISM even at very low densities. Previous studies have shown that when the bubble comes into pressure balance with the ISM, it stops expanding \citep[e.g.][]{mckee77,cioffi88}, leaving a hot, low-density cavity which is later encroached upon by the surrounding gas. However, for very-low density gas \citep[$\nhtot\lesssim 0.01\rm\, cm^{-3}$;][]{cioffi88}, the pressure balance is reached when the bubble is still in the energy-conserving phase and, in these cases, the coupling scheme can fail, injecting too much momentum. To prevent such a behaviour, we limit the momentum coupling to the neighbours with $\rho_j> \rho_{\rm cr}= 0.005 m_{\rm H}\rm\, cm^{-3}$, \footnote{This value corresponds to Eq.~(4.8) in \citet{cioffi88}, assuming $\rho_{\rm gas} = n_{\rm H_{tot}} m_{\rm H}/X$, with $X=0.76$ the cosmic hydrogen fraction.} and for which the distance of the effective face shared with the star $r_{\rm eff}< R_{\rm max}$, where $R_{\rm max}=600\rm\, pc$ is the merging radius for $\rho_{\rm g} =\rho_{\rm cr}$ and $T=10^4$~K \citep{cioffi88},\footnote{The value for $R_{\rm max}$ has been derived from Eqs.~(3.33a) and (4.4b) in \citet{cioffi88} for solar-metallicity gas, $T=10^4$~K, and $\beta=1$, where $\beta$ marks the time at which the merging begins. Nevertheless, the exact value for $R_{\rm max}$ does not significantly affect the results (see Appendix \ref{app:limiter}).} and $r_{\rm eff}=\max\{|\vecdist| - L_j,0\}$, with $L_j = [4\upi/(3N_{\rm ngb})]^{1/3} h_j$ the gas cell effective size.\footnote{With our default choice for the kernel, i.e. the cubic spline, $L_j \approx 0.5 h_j$.} For all the neighbours not fulfilling these criteria, instead, we assume that the momentum $\Delta \mathbf{p}'_j$ has been already converted into turbulent motion, and has finally decayed into thermal energy, which is added to $\Delta U$ as $\Delta E_{{\rm turb},j}= |\Delta\mathbf{p}'_{j}|^2/(2m^{\rm new}_j)$. \footnote{In this case, we are typically in the momentum conserving regime, where a significant part of the energy has already been lost. As a consequence, the addition of thermal energy to the gas is almost ineffective, and does not change the gas temperature significantly.} This choice, compared to that in H18, always guarantees the coupling of the SN ejecta and metal masses even at very low resolution, whereas the default implementation, by only selecting the particles within 2~kpc, can result in missing injection in some extreme cases.

\section{Basic test of the scheme: the single SN case}
\label{sec:test}

\begin{figure}
\includegraphics[width=\columnwidth]{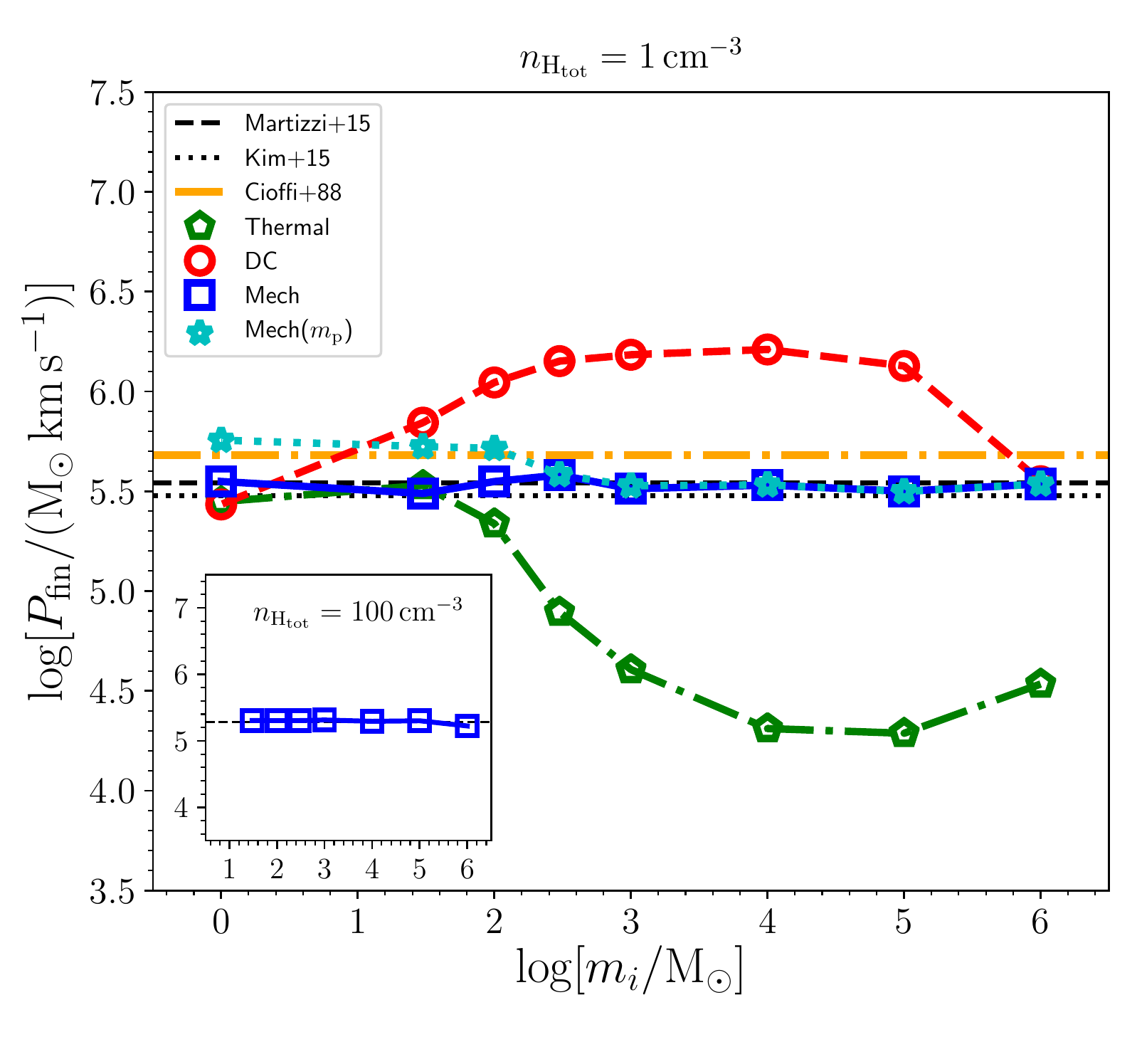}
\caption{Convergence test of the mechanical SN feedback model, via a single SN explosion in a pseudo-homogeneous medium (as in H18) with $\nhtot=1\rm\, cm^{-3}$ (main panel) and $\nhtot=100\rm\, cm^{-3}$ (inset). We show the terminal momentum as a function of mass resolution ($m_i$). The black dashed line corresponds to the terminal momentum as estimated using Eq.~\eqref{eqn:ethpr}, the black dotted one to the terminal momentum from \citet{kimostriker15}, and the orange dot-dashed one to the results by \citet{cioffi88} employed in H18. The mechanical feedback case is shown as a solid blue line (and the blue squares). As a comparison, we also show with a green dot-dashed line (and the green pentagons) the fully thermal case, and with the red dashed line (and the red circles) the delayed-cooling case. Our model can accurately reproduce the terminal momentum independent of the resolution, both at the typical densities of SF sites (in the inset) and in the ISM (main panel). If we instead assume that the swept mass is always equal to the total particle mass (dotted cyan line with the cyan stars), then the terminal momentum is correct only at low resolution, when the residual thermal energy is negligible, but it increases at higher resolution, because of an overestimation of the coupled mass, hence of the total energy injected.} 
\label{fig:pfinal}
\end{figure}

We now test the ability of the mechanical feedback model to reproduce the terminal momentum of the SN bubble expansion at different mass resolutions. The setup of these simulations is the same as in H18, i.e. an homogeneous box of arbitrary size, where particles of mass $m_i$ (i.e. the mass resolution) are randomly distributed using a Monte Carlo sampling. This is a better choice compared to a regular grid, because it mimics the irregular distribution in less idealised simulations, and also highlights how small numerical density fluctuations do not affect our conclusions. Self-gravity is switched off, and we include radiative cooling and non-equilibrium chemistry for 9 primordial species (H,H$^+$,H$^-$,He,He$^+$,He$^++$,H$_2$,H$_2^+$, and e$^-$) using the model described in \citet{lupi18a}. In particular, we consider an extragalactic ultraviolet (UV) background following \citep{haardt12}, photoheating, H$_2$ UV pumping, Compton cooling, photoelectric heating, atomic cooling, H$_2$ cooling, and chemical heating and cooling. Metal cooling is computed from look-up tables obtained with \textsc{cloudy} \citep{ferland13} and tabulated by \citet{shen13}. To model a SN explosion, we inject $10^{51}$~erg, together with $10\,\msun$ of mass at $t=0$ (we do not consider pre-heating of the gas by stellar radiation), but we do not inject metals, to prevent the change in metallicity, which is kept solar, hence the change in the cooling rates. After the SN bubble has reached the momentum-conserving phase, we measure the radial momentum of the expanding gas, and we compare it with the analytic formulae by \citet{cioffi88}, \citet{kim14}, and \citet{martizzi15}. 

The results are shown in Fig.~\ref{fig:pfinal} at $t=1.0$~Myr for two different densities, i.e. $\nhtot=1$ and $100\rm\, cm^{-3}$ (in the inset). As a comparison, for the $1\rm\, cm^{-3}$ case only, we also report the fully thermal model (as a green dot-dashed line with the green pentagons), where we inject $10^{51}$~ erg/SN as thermal energy only, and the delayed-cooling model (as a red dashed line with the red circles), where, in addition, we shut off radiative cooling for the survival time of the blast-wave $t_{\rm max} = 10^{6.85}E_{51}^{0.32} \nhtot^{0.34} \tilde{P}_{04}^{-0.70}$~yr, where $\tilde{P}_{04}=10^{-4} P_0 k_{\rm B}^{-1}$, with $P_0$ the ambient pressure and $k_{\rm B}$ the Boltzmann constant \citep{stinson06}.
The mechanical feedback implementation is well in agreement with the analytic formula at all mass resolutions, up to $m_j=10^6\,\msun$. Nevertheless, at this very low resolution, the ISM structure, as well as single SN events, are not properly resolved, hence specifically designed sub-grid models should be employed in this case \citep[see, e.g.][]{springel03,vogelsberger14,schaye15}. The thermal injection model, instead, is very sensitive to resolution, and the terminal momentum drops when the cooling radius is not resolved, because of the overcooling problem. On the other hand, the delayed-cooling model tends to overestimate the terminal momentum at low resolution (approaching the energy-conserving solution). 
At higher density, the mechanical feedback model still reproduces the terminal momentum very accurately over the entire range of mass resolutions we investigated. This also demonstrates that the model works properly in different density regimes, both at the high densities typical of the sites of SF and at those of the ISM. Unlike the original model by H18, the terminal momentum we obtain is lower, in agreement with the high-resolution simulations by \citet{kimostriker15} and \citet{martizzi15}. In addition, we also test how the results would be affected by always taking the $j$-particle mass as swept mass (cyan dotted line with cyan stars) like in H18. It is interesting to notice that, in this case, at high resolution the total energy injected is larger than $E_{\rm SN}$, because of the overestimation of the radial momentum, and the terminal momentum converges to that of \citet{cioffi88}. At low resolution, when the thermal contribution is negligible and we are well in the momentum-conserving regime, the exact choice of the swept-up mass is not crucial, and the resulting terminal momentum only depends on the value assumed for it.

\section{The isolated galaxy case: mechanical versus delayed-cooling supernova feedback}
\label{sec:setup}
We briefly summarise here the sub-grid model we employ in our isolated galaxy simulations. The chemical network and cooling/heating processes are the same as in \citet{lupi18a}, already described for the single SN test, and include non-equilibrium chemistry for 9 primordial species, with H$_2$ formation via H$^-$ associative detachment and on dust \citep[see][for details]{bovino16}, an extragalactic ultraviolet (UV) background following \citet{haardt12}, and metal cooling look-up tables obtained with Cloudy \citep{ferland13} and tabulated by \citet{shen13}.  Stellar radiation is implemented by collecting all the stellar sources in the gravity tree, and using the total luminosity in the tree nodes when the particles are far away, in a similar fashion to the gravity computation. This approach corresponds to model {\it (b)} in \citet{lupi18a}, and showed the best agreement with on-the-fly radiative transfer simulations.

Compared to our previous study, we slightly modify here the SF prescription, to take into account the pressure support when we approach transonic/subsonic regime, following a similar approach to \citet{semenov17}. Our SF prescription is based on the theoretical studies of turbulent magnetised clouds by \citet{federrath12}. The main assumption in the model, calibrated against simulations and observations, is that the gas follows a Log-Normal density distribution, described by two main quantities, the mean density $\rho_0$ and the width of the corresponding gaussian distribution $\sigma_s = \ln(1+b^2\mach^2)$, where $b$ accounts for the ratio between solenoidal and compressive modes. In our case, we assume a statistical mixture of the two modes, which gives $b=0.4$. By averaging the SF rate over the density distribution, we get the net SF efficiency $\varepsilon$ as
\begin{equation}
\varepsilon=\frac{\varepsilon_\star}{2\phi_{\rm t}}\exp\left({\frac{3}{8}\sigma_s^2}\right)\left[1+{\rm erf}\left({\frac{\sigma_s^2-s_{\rm crit}}{\sqrt{2\sigma_s^2}}}\right)\right],
\label{eq:sfeff}
\end{equation}
where $\varepsilon_\star=0.5$ is the local SF efficiency to match observations \citep{heiderman10}, $1/\phi_{\rm t}=0.49$ is a fudge factor to take into account the uncertainty in the free-fall time-scale, and $s_{\rm crit}= \ln{(\rho_{\rm crit}/\rho_0)}$, with $\rho_{\rm crit}$ the minimum density for SF within the cloud.
In this variant of the prescription, we slightly change our implementation as follows:
\begin{itemize}
\item we replace the velocity dispersion $\sigma_{\rm v}$ in the definition of the virial parameter with the effective energy support against gravitational collapse, i.e. $\sigma_{\rm eff} = \sqrt{\sigma_{\rm v}^2 + c_{\rm s}^2}$, where $c_{\rm s}$ is the sound speed, obtaining
\begin{equation}
\alpha_{\rm vir} =  \frac{5\sigma_{\rm eff}^2L}{6\grav M_{\rm cloud}} = \frac{5(\sigma_{\rm v}^2+c_{\rm s}^2)L}{6\grav M_{\rm cloud}},
\end{equation}
where $L$ is the cloud diameter (assumed to be the grid-equivalent cell size) and $M_{\rm cloud}$ is the particle mass. Assuming the cloud is spherical with $M_{\rm cloud} = 4\upi/3\rho_{\rm gas}(L/2)^3$, we can rewrite $\alpha_{\rm vir}$ as
\begin{equation}
\alpha_{\rm vir} = \frac{5[\|\nabla\otimes\mathbf{v}\|^2+(c_{\rm s}/L)^2]}{\upi\grav \rho_{\rm gas}},
\end{equation}
where $\sigma_{\rm v} = L\|\nabla\otimes\mathbf{v}\|$.
Thanks to the addition of the sound speed to the virial parameter, the SF is self-consistently reduced when we approach a Mach number $\mach=2$, a regime where the sound speed can play a non-negligible role in counteracting the gravitational collapse.
\item in \citet{lupi18a}, the velocity dispersion was computed using the slope-limited velocity gradients. While the slope-limiting procedure is fundamental for the stability of the Riemann solver, it is undesirable here, since it can artificially produce zero velocity dispersion, preventing some gas particles from forming stars, and creating long-lived dense clumps.\footnote{This effect was not visible in the runs by \citet{lupi18a}, since the delayed-cooling model was able to heat up the gas, dissolving these clumps.} In our new variant, we use the `real' gradient, before applying the slope-limiting procedure.
\end{itemize}

As for the SN feedback, we employ here two different models: the new mechanical feedback model we introduced in the previous section, and the delayed-cooling prescription already described in \citet{lupi18a}.
The stellar evolution model, with the different feedback mechanisms considered (type II/Ia SNe and mass losses from low-mass stars) is the same as in our previous study, with the exception of the SN event distribution. While in the previous study SNe were modelled via continuous injection, here we distribute energy, mass, and metals in discrete events. At every time-step, we compute the number of SNe $x_{\rm SN}$ expected to explode according to a Chabrier initial mass function \citep[IMF;][]{chabrier03}, with the stellar lifetimes from \citet{hurley00}. Then, we `discretize' $x_{\rm SN}$ as $N_{\rm SN} = \left\lfloor{x_{\rm SN}}\right\rfloor + \theta(\{x_{\rm SN}\}-p)$, where $\theta$ is the Heavyside step function and $p$ is a uniformly generated random number between 0 and 1. For each type II SN event, we release $E_{\rm SN}=10^{51}$~erg of energy and the IMF-averaged ejecta masses, respectively a total ejecta mass $M_{\rm ej}=15.1452-M_{\rm NS}=13.7452\,\msun$, with $M_{\rm NS}=1.4\,\msun$ the mass of the remnant neutron star, an oxygen mass $M_{\rm oxy}=1.2403\,\msun$ and an iron mass $M_{\rm iron}=0.10422\,\msun$. The total metal mass injected per single event corresponds to $M_{\rm Z} = 2.09M_{\rm oxy}+1.06M_{\rm iron}=2.7028\,\msun$. Our prescription results in about 1 SN every $91\,\msun$ formed.
For type Ia SNe, instead, we inject $10^{51}$~erg, $M_{\rm ej}=1.4\,\msun$, $M_{\rm oxy}=0.14\,\msun$, and $M_{\rm iron}=0.63\,\msun$, with a total metal mass injected of $M_{\rm Z}=0.9604\,\msun$. Finally, for low-mass stars, we expect low-velocity winds, hence we neglect the additional momentum of the wind and only distribute the released mass and the initial momentum to the neighbours. During a Hubble time, our prescription is able to recycle 42 per cent of the initial stellar mass \citep[see][and references therein]{kim14}. In order to resolve single SN events in our simulations, we also add a time-step limiter for the stellar particles. For particles younger than 100 Myr, that are dominated by Type II SN events, we limit the time-step to 1/100th of the typical lifetime of a $40\,\msun$ star, which is in the range $3-5\times10^4$~yr. For older particles, instead, we increase this limit to 1/10th of the age of the stellar population.

\subsection{Simulation suite}
\begin{table}
\centering
\caption{Simulation suite for the isolated galaxy. In the first column, we report the name of the run, in the second the minimum gas softening, in the third one the stellar particle softening, and in the last one the dark matter particle softening.}
\begin{tabular*}{\columnwidth}{@{\extracolsep{\fill}} lcccc}
\hline
Run & $m_{\rm gas}$ & $\epsilon_{\rm gas,min}$ & $\epsilon_\star$ & $\epsilon_{\rm DM}$\\
\hline
VLmech & 1.7e5~$\msun$ & 13.0~pc & 25.0~pc & 100~pc\\
Lmech & 8.6e4~$\msun$ & 10.0~pc & 20.0~pc & 80~pc\\
Mmech & 1.7e4~$\msun$ &   4.2~pc & 8.0~pc & 50~pc\\
Hmech & 8.6e3~$\msun$ &   3.0~pc & 6.0~pc & 40~pc\\
VHmech & 1.7e3~$\msun$ &   1.7~pc & 3.5~pc & 23~pc\vspace{0.1cm}
\\
Hdc & 8.6e3~$\msun$ & 3.0~pc & 6.0~pc & 40~pc\\

\hline
\end{tabular*}
\label{tab:suite}
\end{table}

To compare the two SN feedback models, we evolve a Milky Way-like galaxy at $z=0.1$ for 500~Myr in isolation. The initial conditions (ICs) are those of the AGORA collaboration \citep{kim16}. We initially relax the ICs for 500~Myr adiabatically, to avoid the initial numerical fragmentation due to the initial density fluctuations, and then enable the sub-grid prescriptions implemented, evolving the galaxy for additional 500~Myr. We run five simulations with our new mechanical feedback model, at different mass/spatial resolutions, and one simulation at high resolution with the delayed-cooling model. At the five resolutions, the gas is sampled with $ 5\times 10^4$,$10^5$,$5\times 10^5$, $10^6$, and $5\times 10^6$ particles, respectively. The full suite of simulations is reported in Table~\ref{tab:suite}.

\begin{figure}
\centering
\includegraphics[width=\columnwidth]{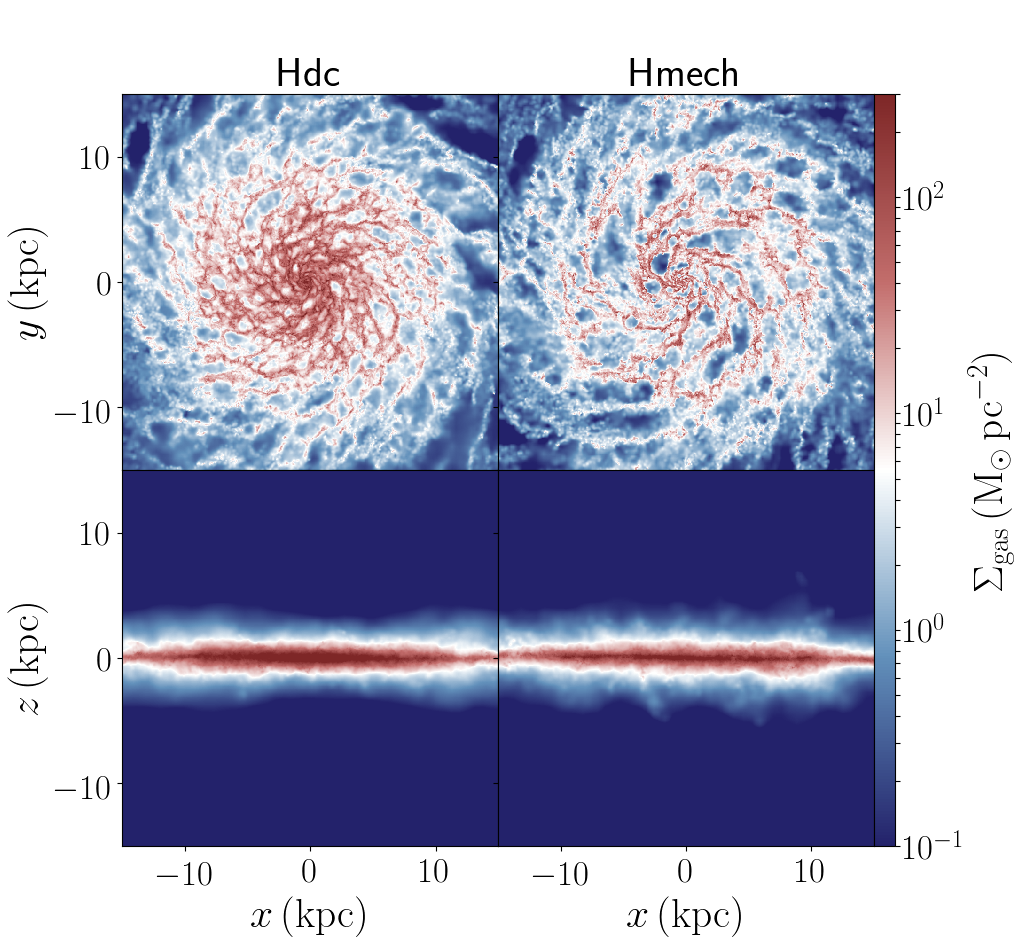}
\caption{Face-on (top panels) and edge-on (bottom panels) views of the galaxy gas surface density at $t=500$~Myr for Hmech (right-hand panels) and Hdc (left-hand panels). Because of the lower SFR in Hdc, more gas has remained in the galaxy, producing an overall higher gas density. Nevertheless, the density contrast is larger in Hmech (face-on view), where we observe low-density cavities surrounding high-density SF regions, whereas Hdc exhibits a smoother density distribution. Observed edge-on, the dense disc in Hmech is slightly thinner than that in Hdc, although the low-density gas above and below the disc is very similar in the two runs.}
\label{fig:dvsmmap}
\end{figure}

\section{Results}
\label{sec:results}
\subsection{Delayed-cooling versus mechanical feedback}
We discuss now how the galaxy ISM and the SF evolve with the two SN feedback models, by comparing the fiducial simulations, Hmech and Hdc, at the end of the run ($t=500$~Myr).
\subsubsection{Gas distribution}

\begin{figure}
\centering
\includegraphics[width=0.88\columnwidth]{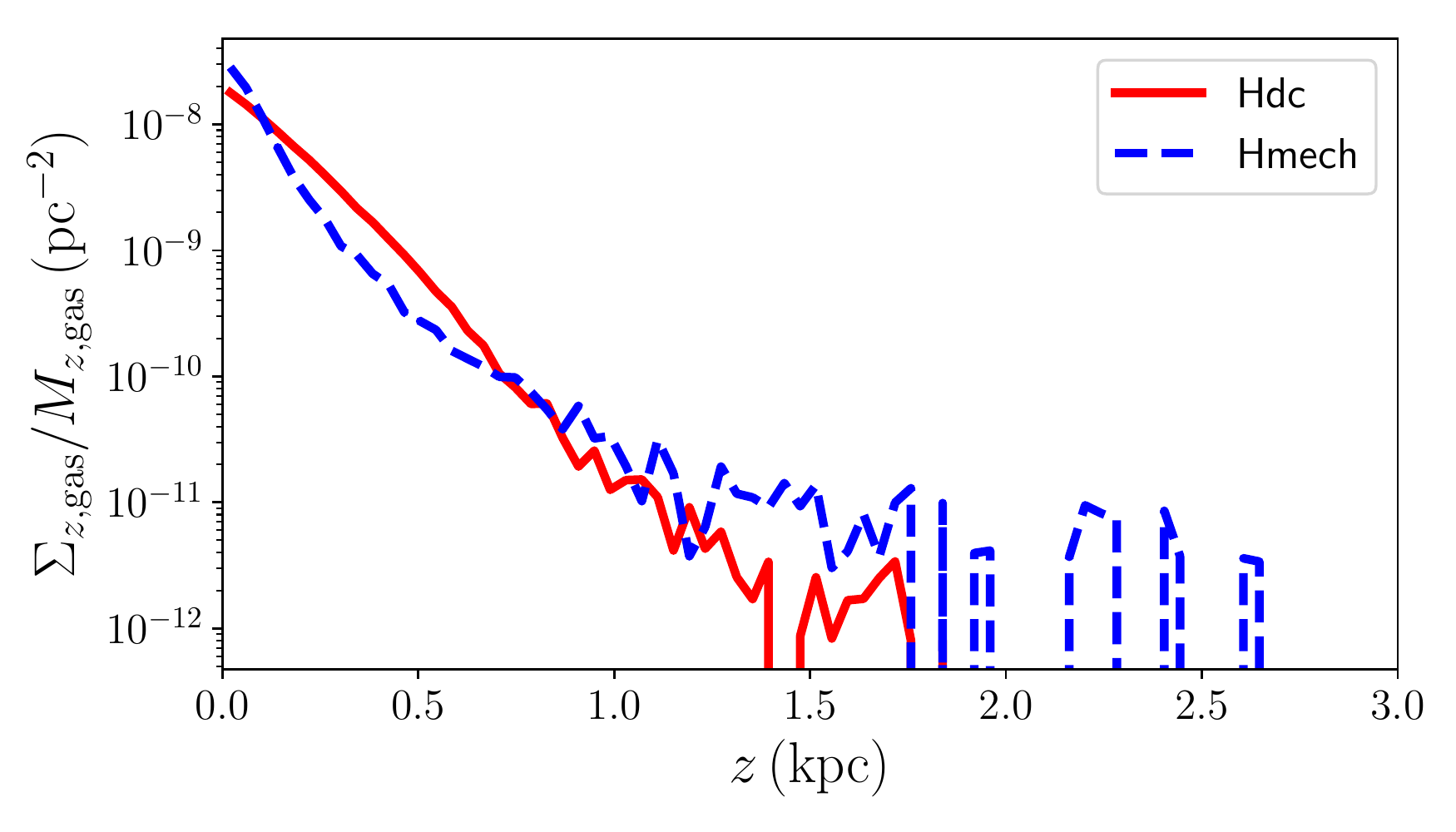}
\caption{Vertical surface density profile at $t=500$~Myr for Hdc (red solid) and Hmech (blue dashed), computed within a cylinder of 2~kpc radius around the galaxy centre, normalised to the total gas mass in the cylinder. The pressurising effect of the delayed-cooling prescription makes the disc in Hdc slightly thicker than in Hmech. In particular, if we assume the scale height $H$ of the disc being the distance where the density is half that at the peak, we get $H\sim 0.15$~kpc for Hmech and $H\sim 0.2$~kpc for Hdc. Although the actual vertical profile changes with time because of SNe, the intrinsic difference between the two models is preserved.}
\label{fig:dvsmsigma}
\end{figure}
\begin{figure}
\centering
\includegraphics[width=0.88\columnwidth]{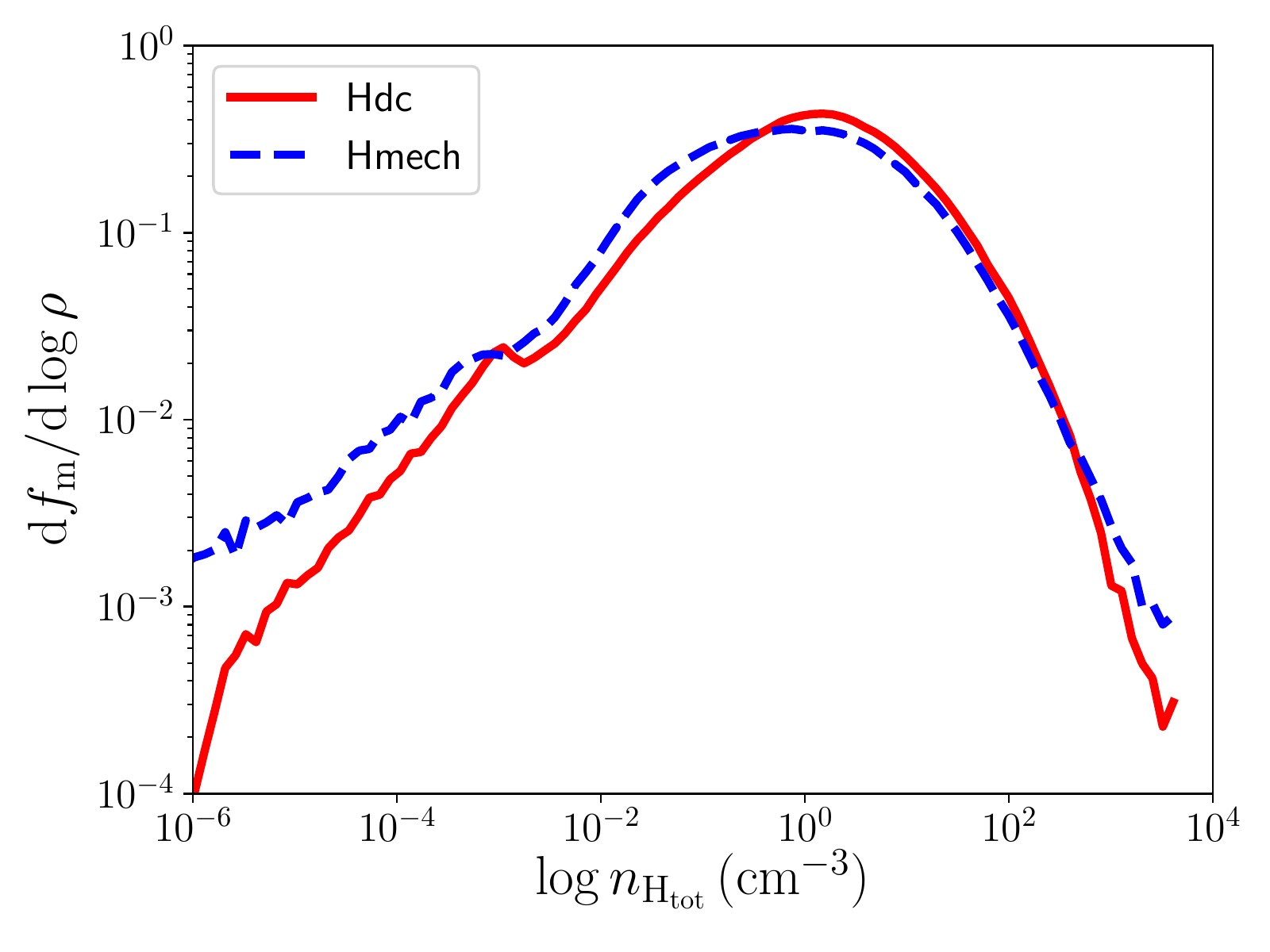}
\caption{Gas mass distribution as a function of total hydrogen nuclei density for Hmech (as a blue dashed line) and Hdc (as a red solid line) at $t=500$~Myr, normalised to the total gas mass available. The two profiles show a similar distribution, but with a mild offset to lower densities in Hmech compared to Hdc, and a much more important low-density tail (due to the gas outflows powered by SNe). Although not shown, we repeated the comparison at different times, and the results do not change significantly.}
\label{fig:dvsmphase}
\end{figure}

First, we show in Fig.~\ref{fig:dvsmmap} a qualitative comparison of the gas surface density of the galaxy, face-on and edge-on. The Hdc run is reported in the left-hand panels, whereas Hmech is in the right-hand ones. Hdc shows a higher gas density than Hmech in the central 10~kpc, because of the higher SFR in Hmech which consumed a larger fraction of the available gas (see Fig.~\ref{fig:dvsmsfr}). Nonetheless, the pressurising effect of the cooling shut-off makes the gas distribution in Hdc slightly smoother than in Hmech, where very high-density SF sites are instead surrounded by the low-density cavities carved by SNe. The edge-on view shows that the dense region of the disc in Hdc is slightly thicker than in Hmech, because of the pressurising effect of the delayed-cooling SN feedback, which keeps the gas hotter. This can be clearly seen in Fig.~\ref{fig:dvsmsigma}, where we show the vertical surface density profile, obtained as the surface density of the gas within a cylinder of 2~kpc radius at the centre of the galaxy, binned along the $z$ axis. To avoid the confusion coming from the different amount of available gas, the profiles are normalised to the total gas mass $M_{\rm z,gas}$ within the cylinder.

\begin{figure}
\centering
\includegraphics[width=\columnwidth,trim={1.3cm 0cm 0.5cm 0cm},clip]{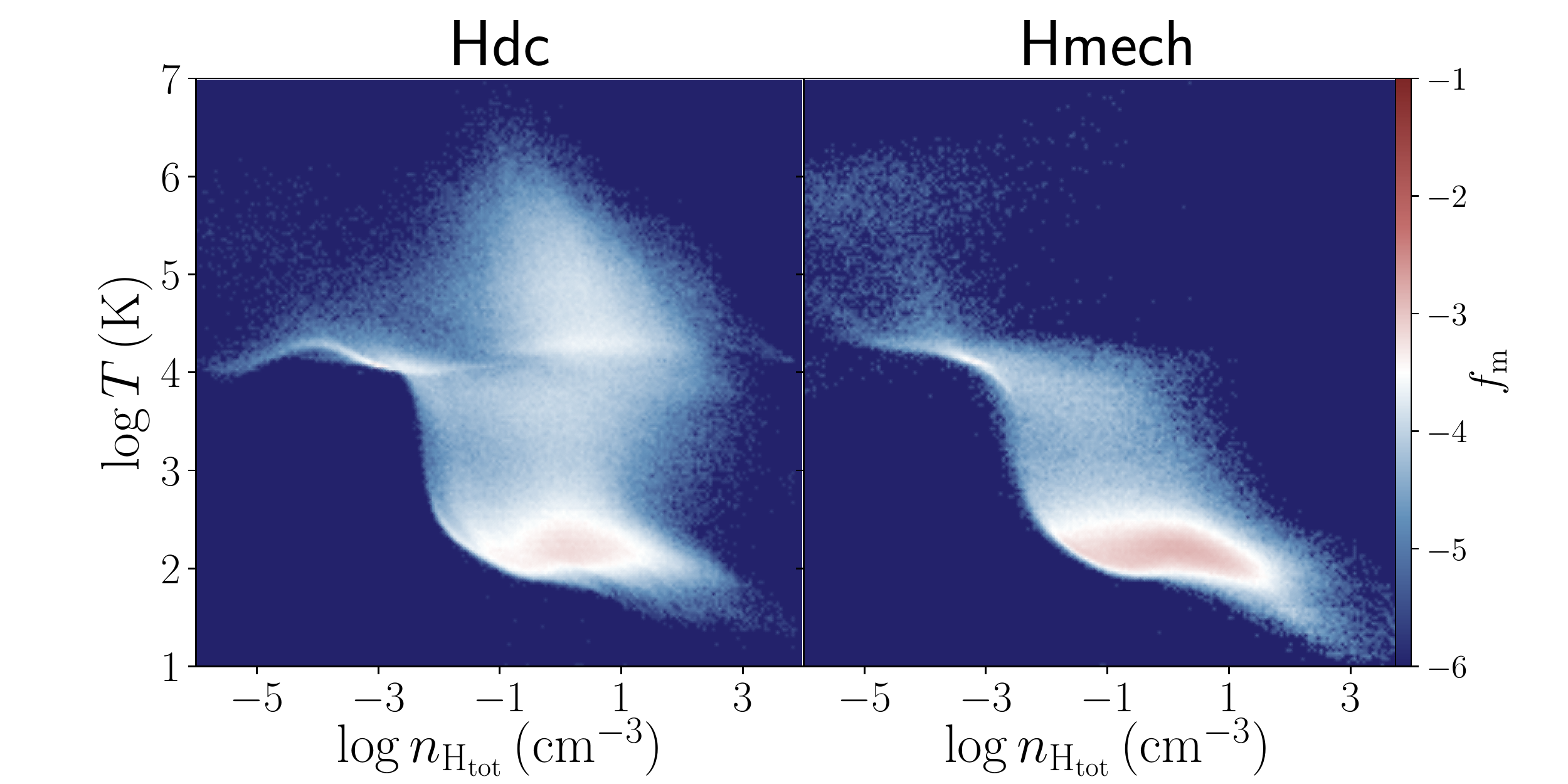}
\caption{Density--temperature diagram for Hmech (as a blue dashed line) and Hdc (as a red solid line)  at $t=500$~Myr. At densities typical of the ISM and SF regions, Hdc can keep the gas hot (above $10^4$~K), spreading the gas over 4--5 orders of magnitude in temperature, whereas Hmech allows it to cool down, as it should naturally do.}
\label{fig:dvsmrhoT}
\end{figure}

\begin{figure}
\includegraphics[width=\columnwidth,trim={1.3cm 0 0.5cm 0},clip]{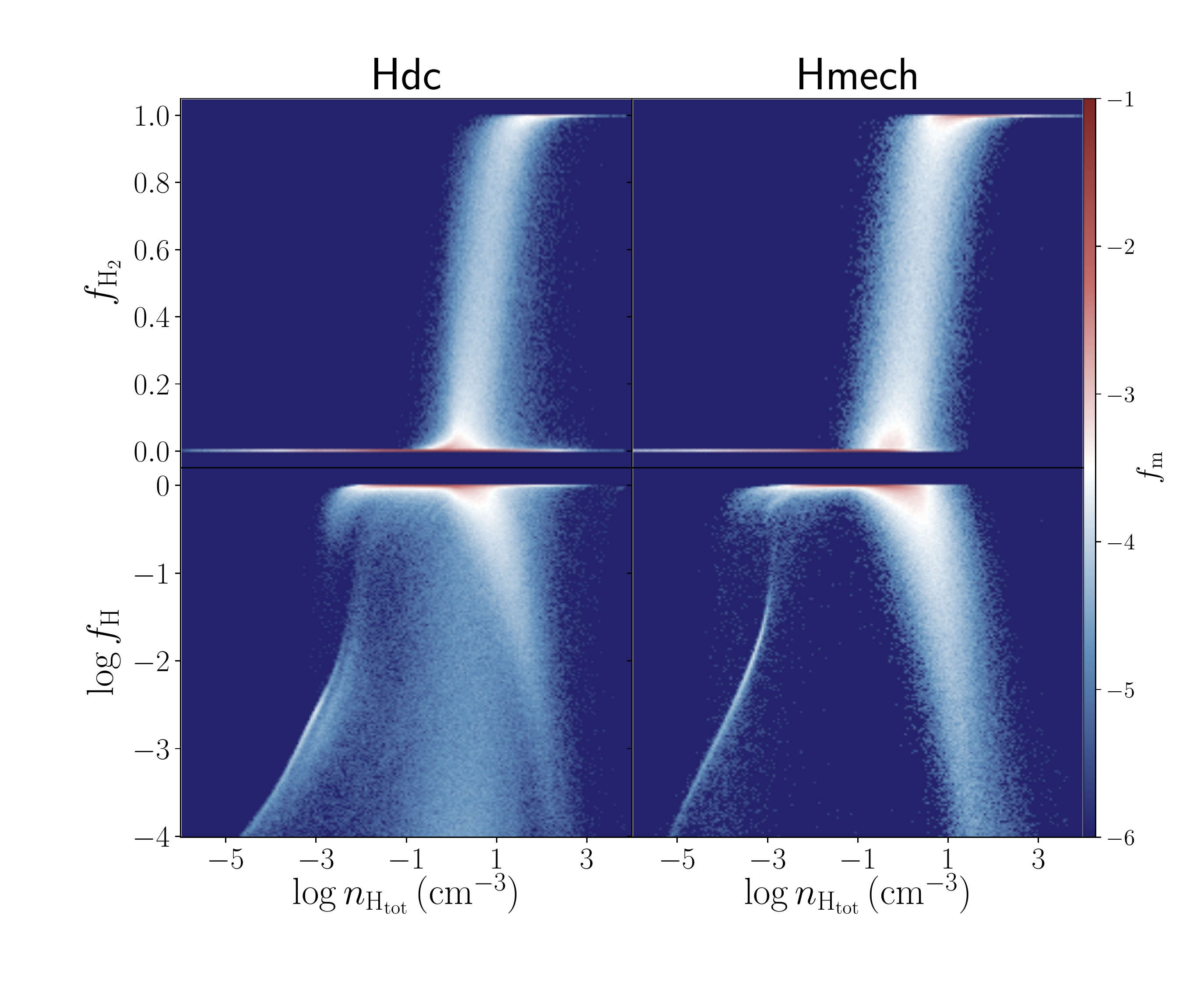}
\caption{Atomic and molecular hydrogen abundances for Hdc (left-hand panels) and Hmech (right-hand panels), as a function of $\nhtot$. The colour map corresponds to the gas mass fraction $f_{\rm m}$. Hmech exhibits a larger amount of H$_2$ (2.5 times the mass in Hdc), because of the typically lower temperature of the gas compared to Hdc, where gas is more efficiently heated up by SNe, and the high density tail of atomic/ionised hydrogen in Hdc is not visible anymore in Hmech. As for the H abundance, instead, Hmech shows almost no H$^+$ for $0.01\rm\, cm^{-3}\lesssim \nhtot \lesssim 1.0\rm\, cm^{-3}$, while Hdc, thanks to the cooling shut-off, keeps part of this gas hot and ionised. }
\label{fig:dvsmabund}
\end{figure}

\begin{figure}
\centering
\large 
Hdc\\
\vspace{0.2cm}
\includegraphics[width=0.8\columnwidth,trim={1.3cm 0 0.5cm 0},clip]{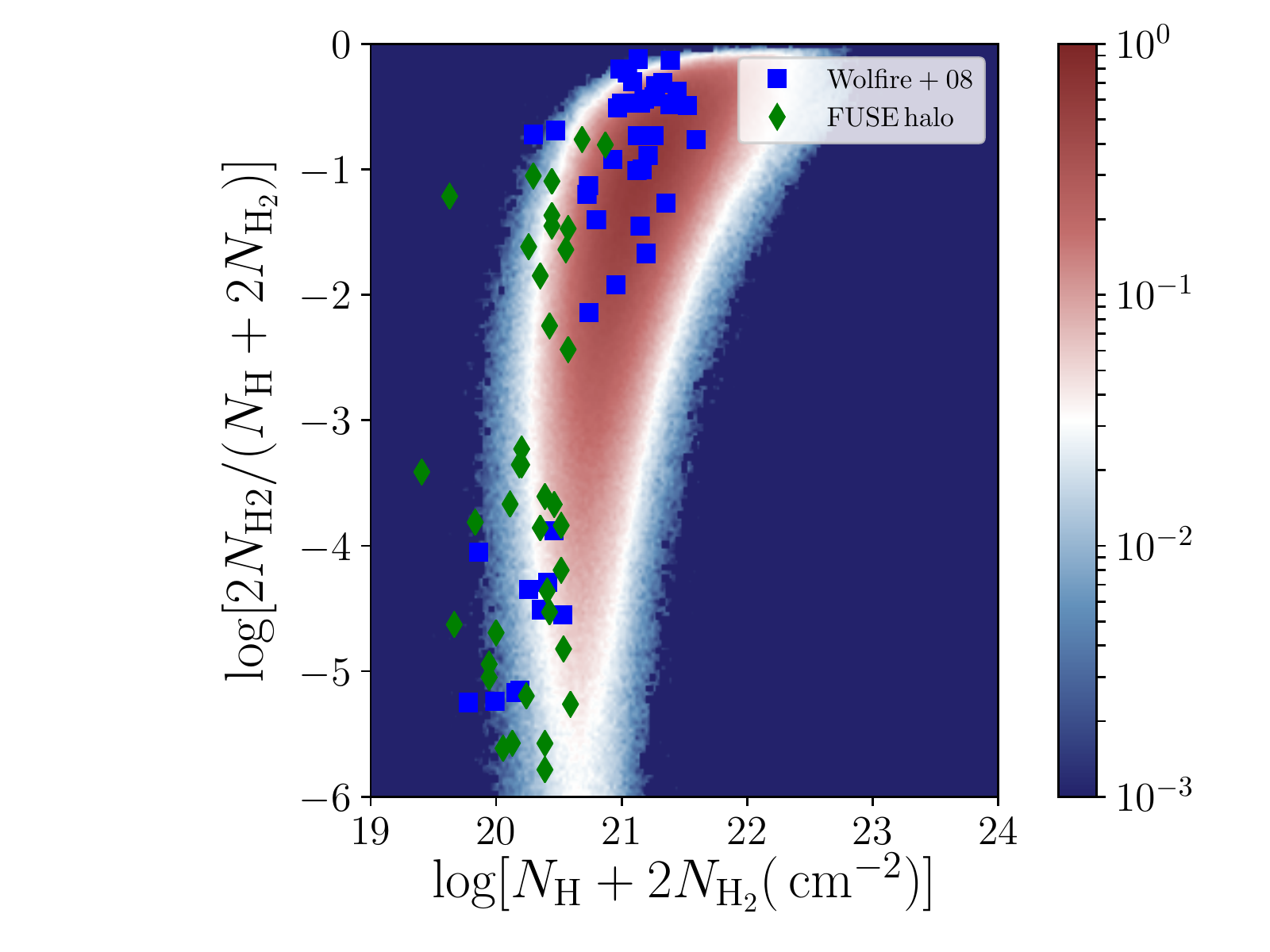}\\
Hmech\\
\vspace{0.2cm}
\includegraphics[width=0.8\columnwidth,trim={1.3cm 0 0.5cm 0},clip]{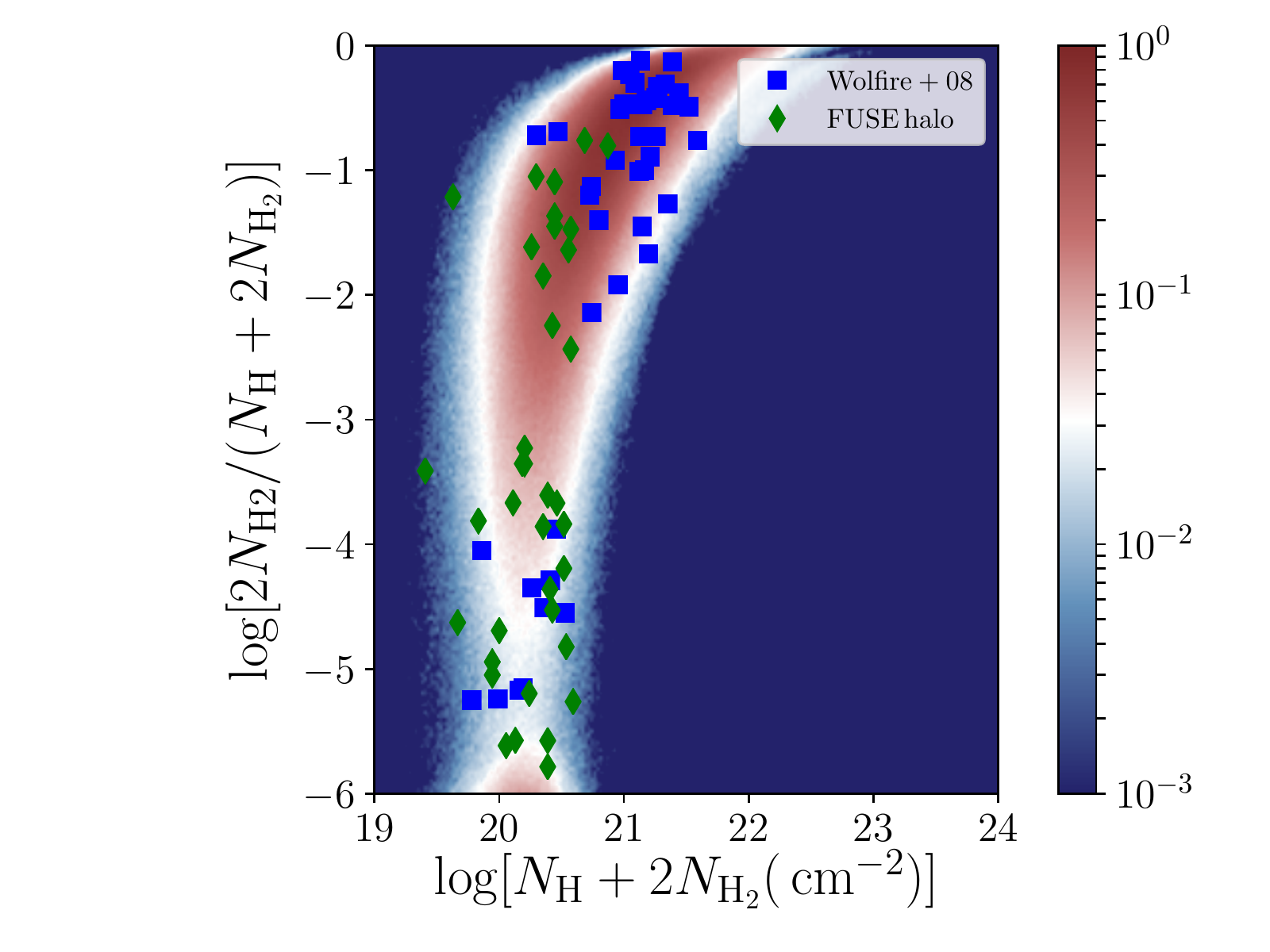}\\
\caption{H$_2$ column density fraction in Hdc (top panel) and Hmech (bottom panel) at t = 500 Myr, compared with local observations of molecular clouds in the Milky Way disc and halo, and in the Large Magellanic Cloud. Our simulation results overlap very well with observations across the entire available column density range, with Hmech showing a remarkable agreement. Thanks to the typically lower gas temperatures (a crucial difference with Hdc, where cooling shut-off keeps gas hot above $\nhtot \sim 0.1-1\rm\, cm^{-3}$, artificially suppressing H$_2$ formation), H$_2$ is significantly more abundant in Hmech, shifting the distribution upwards, allowing it to more easily reach the fully molecular regime.}
\label{fig:dvsmcolH2}
\end{figure}

Fig.~\ref{fig:dvsmphase} shows the gas mass distribution within the galaxy. We plot the gas mass fraction $f_{\rm m} = m_{\rm gas}/M_{\rm gas}$ as a function of $\nhtot$, where $M_{\rm gas}$ is the total available gas mass in the galaxy. We binned the gas particles in logarithmically spaced density bins 0.1 decades wide. The gas follows a generally similar distribution, described by a Log-Normal profile at high density and a power-law tail at low density. Nevertheless, some differences can be clearly observed. The amount of gas at $\nhtot\gsim 1\rm\, cm^{-3}$ is slightly lower in Hmech than in Hdc, because of the stronger dynamical effect of the mechanical SN feedback which more effectively kicks the particles above and below the disc. This gas is indeed found at much lower densities ($\nhtot<10^{-3}\rm\, cm^{-3}$), in the power-law tail, where we observe differences of up to one order of magnitude between the two runs.

In Fig.~\ref{fig:dvsmrhoT}, we report the density--temperature diagram of the gas for the same two runs (Hmech in blue and Hdc in red), where the colour map corresponds to the mass fraction $f_{\rm m}$. In Hdc, SNe keep the gas hot, even at densities where it should normally quickly cool down, spreading the gas temperature distribution over 4--5 orders of magnitude. In Hmech, instead, the gas is pushed away by SNe (when the cooling radius is not resolved), getting warmer as it moves to lower densities where the optical depth is lower and it can be heated up and ionised by local (and extragalactic) stellar radiation. This results in a large fraction of gas remaining cold (below $10^4$~K) at all densities above $\nhtot=10^{-2}\rm\, cm^{-3}$, where the gas shielding from the extragalactic UV background becomes important.\footnote{ Compared to Hdc, the lower temperature in Hmech results in a slightly stronger turbulence in the SF sites, with an average velocity dispersion $\langle\sigma_{\rm v}\rangle\sim 25\rm~km~s^{-1}$, an average sound speed $\langle c_{\rm s}\rangle \sim 1.2\rm~km~s^{-1}$, an average $\langle\mach\rangle \sim 22$, and a median $\mach_{\rm median}\sim 18$. The average SF efficiency is therefore $\langle \varepsilon_{\rm SF}\rangle \sim 0.007$ and, rho-weighted, $\langle \varepsilon_{\rm SF}\rangle_{\rm \rho} \sim 0.1$.}

The different temperature distribution of the gas can have important effects on the chemical state of the gas. To assess the dependence of the distribution on the SN feedback model used, we compare in Fig.~\ref{fig:dvsmabund} the mass fraction of molecular hydrogen $x_{\rm H_2}$ (top panels) and atomic hydrogen $x_{\rm H}$ (bottom panels) as a function of $\nhtot$. H$_2$ is 2.5 times more abundant in Hmech (top panels), because of the typically lower temperature of the gas, and the tail of atomic/ionised hydrogen above $10\rm\, cm^{-3}$ observed in Hdc (top-left panel), due to the warm/hot gas heated up by SNe, has disappeared in Hmech (top-right panel). As for H, instead, the density interval $0.01\rm\, cm^{-3}\lesssim \nhtot \lesssim 1\rm\, cm^{-3}$ is completely devoid of ionised gas (corresponding to $f_{\rm H}\ll 0.1$) in Hmech, unlike in Hdc.

Finally, we compare the H$_2$ column density with observations of the Milky Way disc and halo (solar metallicity). We compute the column density of both atomic and molecular hydrogen, respectively N$_{\rm H}$ and N$_{\rm H_2}$, in a cylinder of 20~kpc radius and 2~kpc height along 25 randomly distributed lines of sight. We report in Fig.~\ref{fig:dvsmcolH2} the logarithmic H$_2$ column density fraction as a function of the total logarithmic column density. The results are binned in 200 log-spaced bins along both axes, where the colour coding represents the point density in each bin. The blue squares refer to the Milky Way disc data by \citet{wolfire08}, and the green diamonds to the Milky Way halo from the FUSE survey. Both runs agree reasonably well with observations across the entire available data range, with mild differences consistent with those discussed above. In particular, Hmech shows the best agreement, because of the larger fraction of cold gas (hence H$_2$) compared to Hdc, where the intermediate-density gas kept hot by SNe reduces the H$_2$ content. This results in Hmech being able to reach the fully molecular regime, unlike Hdc.

\subsubsection{The star formation history and the Kennicutt--Schmidt relation}
\label{sec:ks}
\begin{figure}
\centering
\includegraphics[width=0.9\columnwidth]{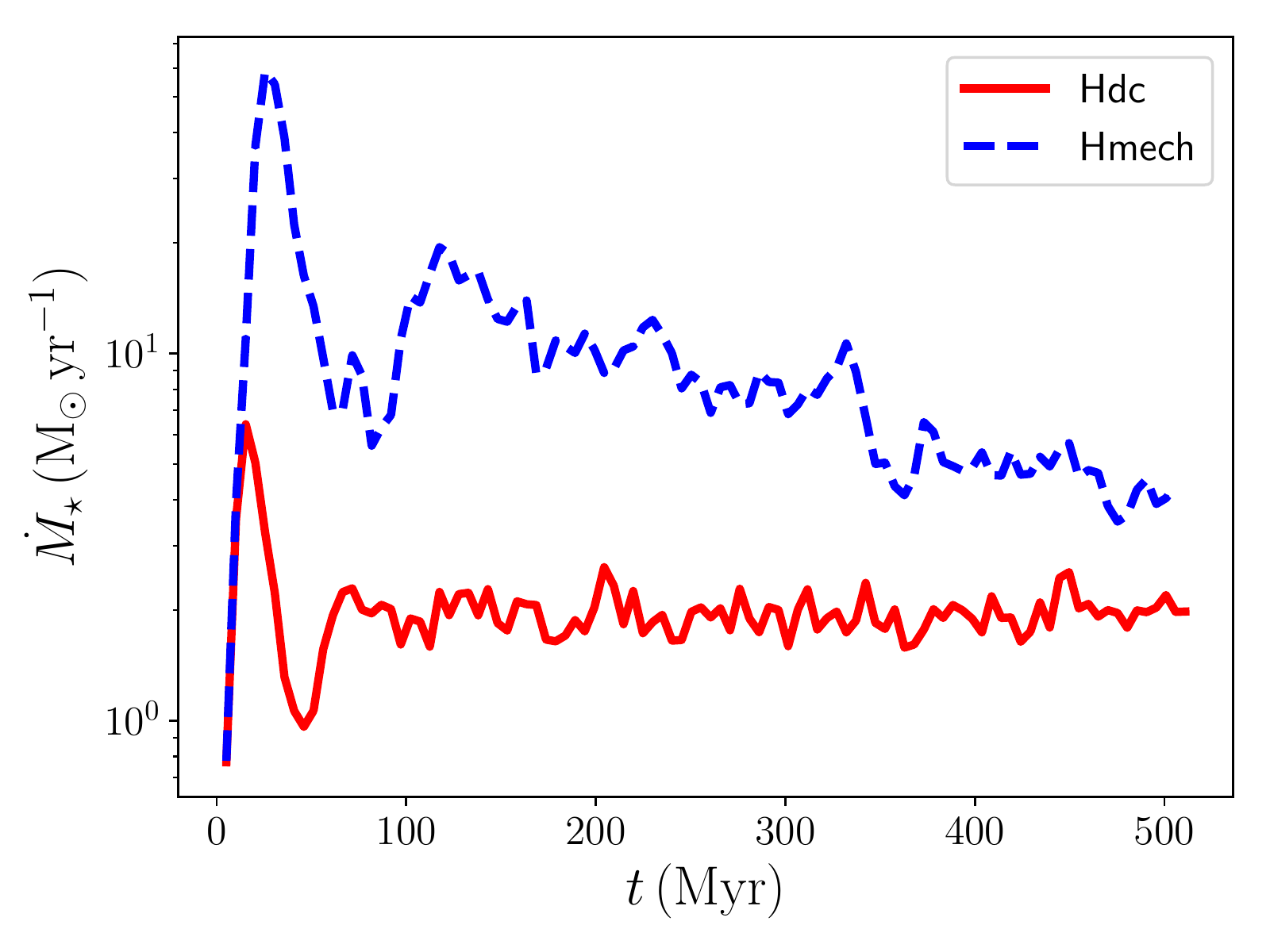}
\caption{SF rate in the Hmech (blue dashed) and Hdc (red solid) runs. After the initial burst of SF, Hdc shows a steady evolution at almost constant SFR of about 1--2~$\msun\rm\, yr^{-1}$, due to the high temperature of the gas affected by SN feedback, which does not form stars for long timescales (a few tens of Myr). In Hmech, on the other hand, SNe do not heat the gas up, but sweep it away, triggering strong shocks with the cold ISM and new cooling that produces new stars. This delays the establishment of a self-regulated equilibrium, and results in more gas being consumed by SF.}
\label{fig:dvsmsfr}
\end{figure}
\begin{figure*}
\centering
\includegraphics[width=0.81\textwidth]{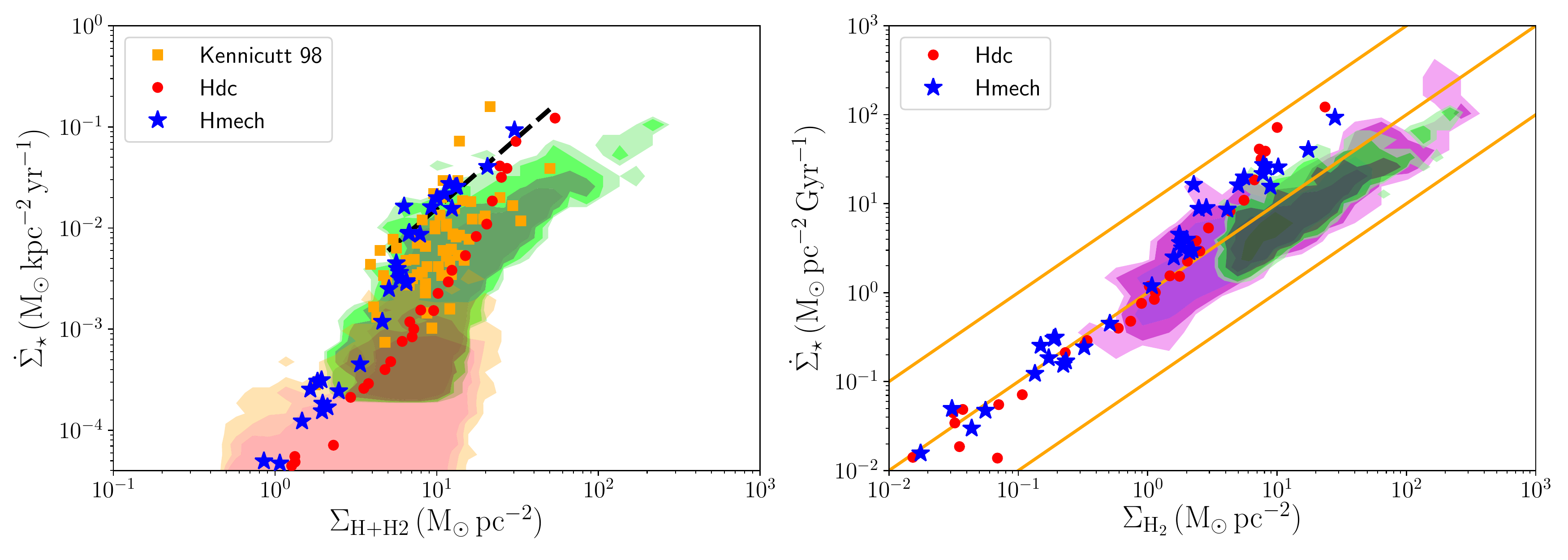}
\caption{KS relation for the Hmech (blue stars) and Hdc (red dots) runs at $t=500$~Myr, in both total gas (H+H$_2$, left-hand panel) and molecular gas (H$_2$, right-hand panel) only, compared with the observational samples by \citet{kennicutt98} (orange stars), \citet{bigiel08,bigiel10}, and \citet{schruba11}. For the simulated data, we show here the average SFR in circular annuli 500~pc wide. Hdc agrees very well with the observed data in both panels, with only a small overshooting at high densities. In the left-hand panel, the higher SFR of Hmech results in the points being closer to the upper boundary of the observational contours, but still consistent with the observed data. Nevertheless, the slope of the relation above $\Sigma_{\rm H+H_2}\sim 10\,\msun\, pc^{-2}$ for Hmech is consistent with the fit by \citet{kennicutt98} (black dashed line). In the right-hand panel, the correlation is better preserved, with Hmech almost overlapping the Hdc results at all densities, but without the overshooting seen in Hdc.}
\label{fig:dvsmks}
\end{figure*}

To assess how effective our mechanical feedback model is in suppressing SF, we show in Fig.~\ref{fig:dvsmsfr} the SF rate in the Hmech (blue dashed) and Hdc (red solid) runs, estimated by sampling the stellar mass formed in bins of 10 Myr. In the first few tens of Myr, radiative cooling removes the vertical pressure support from the disc, making it collapse vertically. This triggers an initial burst of SF of up to a few tens of $\msun\rm\, yr^{-1}$, and a subsequent burst of SN explosions. With the delayed-cooling model, SNe heat up the gas and keep it hot, increasing the pressure support and reducing $\mach$. This turns out in a very quick suppression of the SF down to less than $1\,\msun\rm\, yr^{-1}$. Then, the galaxy settles in a self-regulated state, with a steady SFR in the range $1-2\,\msun\rm\, yr^{-1}$. With the mechanical feedback model, instead, the gas is swept away from the SF sites, but not heated up significantly. Because of the kick imparted by SNe and the low temperature of the gas, strong shocks occur within the ISM, but the subsequent rise in temperature is not enough to avoid rapid cooling of the shocked gas, producing new stars and, subsequently, new SNe. As a consequence, in Hmech, we observe a steady decay instead of an abrupt suppression, with the galaxy taking more time to settle in a self-regulated state. This also results in more gas being consumed by SF than in Hdc, in agreement with the results by \citet{rosdahl17}.

Observations of low-redshift galaxies clearly demonstrated that a correlation exists between the SFR and the gas surface densities in galaxies \citep[KS hereon;][]{schmidt59,kennicutt98}, with a power-law scaling as $\dot{\Sigma}_\star \propto \Sigma_{\rm g}^{1.4}$. Recently, the effort made to collect a large sample of data has allowed us to improve the constraints on this relation, and to show that a tight correlation exists between the SFR and the molecular gas content, whereas almost no correlation exists with the abundance of atomic hydrogen \citep{bigiel08}. In Fig.~\ref{fig:dvsmks}, we compare the ability of the two SN feedback models in reproducing the KS relation, in both total gas (H+H$_2$, left-hand panel) and molecular gas (H$_2$, right-hand panel) only. The filled contours correspond to observations by \citet{bigiel08} (green), \citet{bigiel10} (red), binned in 0.1 decade wide bins on both axes, and \citet{schruba11} (purple), binned in 0.2 decade wide bins on both axes. The different contour levels correspond to the density of data points in the bins, respectively 2, 5, and 10. The orange squares correspond to the observations by \citet{kennicutt98}.
To accurately compare the simulated galaxies with observations, we compute the far ultraviolet (FUV) flux of the stellar sources, according to the age and metallicity of each star particle, and we use the relation in \citet{salim07} ($\dot{M}_\star=0.68\times 10^{-28}L_{\rm FUV}$) to convert the FUV luminosity into a SFR. \footnote{By comparing the FUV-derived SFR with the instantaneous SFR from the gas in the simulation, we find a reasonable agreement, but for the highest gas densities, where the FUV-based method results in a factor of a few lower SFR. This is due to the longer time-scales traced by the FUV luminosity compared to the instantaneous value.} We then estimate the density and the SFR in circular annuli 500 pc wide, with the galaxy seen face-on. For total gas, Hdc (red dots) agrees well with the observed data, whereas Hmech (blue stars) exhibits a higher SFR, with a factor of a few difference relative to Hdc, and is closer to the upper edge of the observed distribution. The black dashed line in the panel, corresponding to the fit by \citet{kennicutt98} shifted upwards by a factor of 2.5, shows that, despite the offset normalisation in Hmech, the slope is in perfect agreement with observations. In the molecular relation (right-hand panel), both models are consistent with the observations, with Hmech and Hdc almost overlapping, except for a mild overshooting in Hdc at high densities. This suggests that, while the KS in total gas can be significantly influenced by the SN feedback model employed, the molecular gas counterpart is less affected by these changes.

\subsubsection{Outflow properties}
\begin{figure*}
\centering
\includegraphics[width=0.725\textwidth]{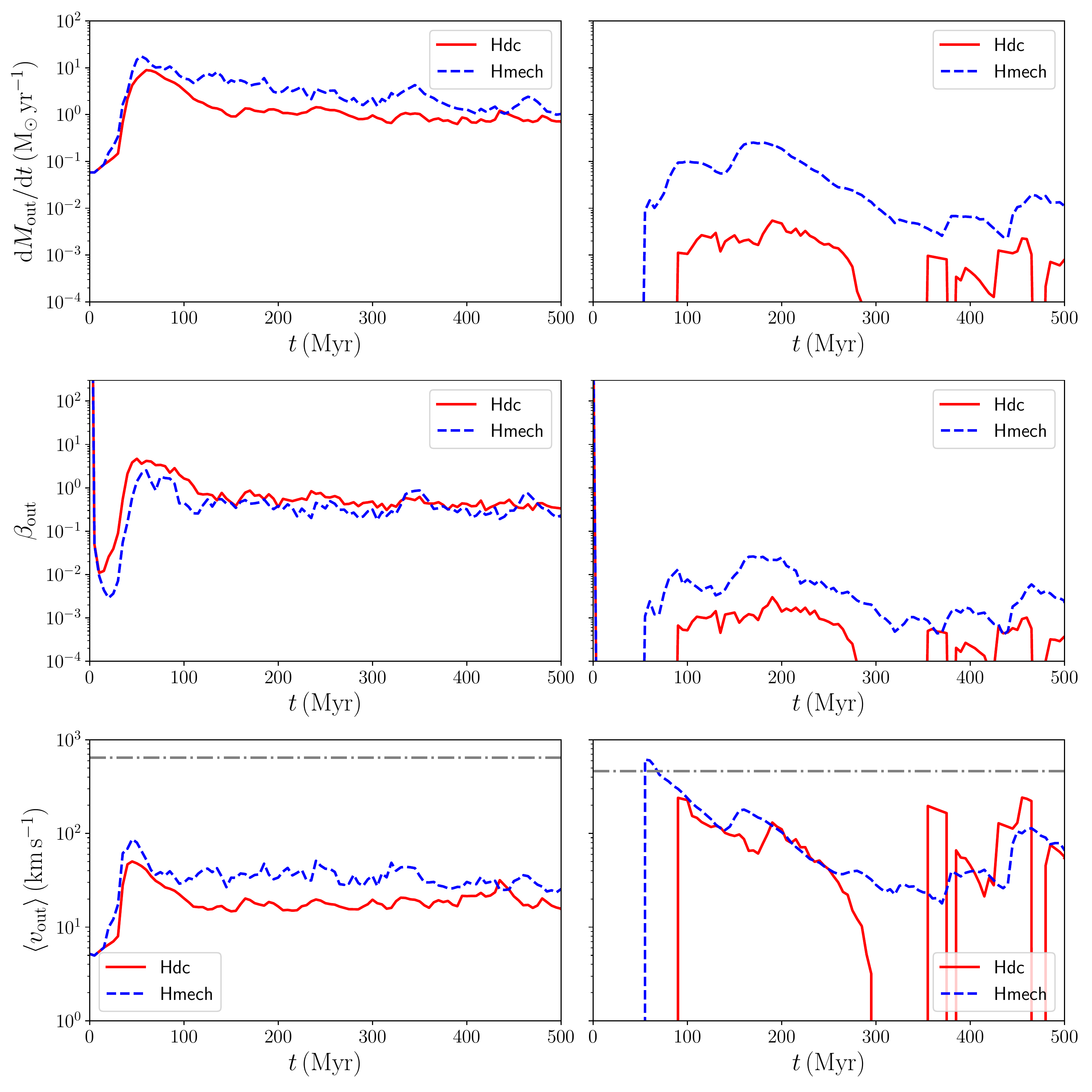}
\caption{Outflow properties for the Hmech (blue dashed) and Hdc (red solid) runs, at 2~kpc (left-hand panels) and 20~kpc (right-hand panels) from the galactic disc plane. In the top panels, we show the mass outflow rate, in the middle panels the $\beta_{\rm out}$ parameter, and in the bottom panels the outflow mean velocity. The dot-dashed lines in the bottom panels correspond to the maximum escape velocity from the disc at $z=2$~kpc (left-hand panel) and $z=20$~kpc (right-hand panel), respectively. Hmech is able to produce stronger outflows relative to Hdc. While this effect could in principle be associated with the higher SFR in Hmech, consistent with the $\beta_{\rm out}$ measured in Hdc at 2~kpc, the results at 20~kpc clearly demonstrate that this is not the case. Indeed, Hdc is very inefficient at producing outflows up to 20~kpc, with mass rates and $\beta_{\rm out}$ roughly one order of magnitude lower than in Hmech.}
\label{fig:dvsmoutflow}
\end{figure*}

Another key aspect to assess the ability of different SN feedback models into regulating SF is the production of outflows. Gas ejected from SNe moves into the halo (in some cases, it can also escape the halo gravitational potential), falling back at later times and triggering a new phase of SF. Galactic outflows are routinely observed in galaxies \citep{steidel10,heckman15}, but many properties as the mass outflow rate and the density/temperature state of the gas are not constrained very well.
In Fig.~\ref{fig:dvsmoutflow}, we compare the outflow properties for the two SN feedback models, i.e. the mass outflow rate, the mass loading factor $\beta_{\rm out}=\dot{M}_{\rm out}/\dot{M}_\star$, and the outflow velocity. 
We measure the outflows at 2~kpc (left-hand panels) and 20~kpc (right-hand panels) from the galactic plane, selecting the gas particles with $v_z\cdot z>0$, with $v_z$ the vertical velocity, in a slab with $\Delta z = 0.2$~kpc and $\Delta z = 5.0$~kpc, respectively. \footnote{This choice guarantees that the poor mass/volume sampling at low resolution does not affect the measure at large distances.} This corresponds to the gross outflow (neglecting any inflowing gas). The mass outflow rate at 2~kpc (top-left panel) is a factor of a few higher in Hmech than in Hdc. At 20~kpc, the difference is even larger, with Hmech being roughly one order of magnitude higher than Hdc. At late times, when the disc has reached a self-regulated equilibrium, the outflow rate is more similar between the two runs, but significantly low ($\dot{M}_{\rm out} \lesssim 10^{-2}\,\msun\rm\, yr^{-1}$). The situation is more complex for $\beta_{\rm out}$, because of the additional dependence on the SFR. At 2~kpc, the larger outflow rates in Hmech compensate the higher SFRs relative to Hdc, resulting in a comparable $\beta_{\rm out}\sim 0.5-1$. At larger distances, instead, Hmech can reach $\beta_{\rm out} \sim 10^{-2}$, in particular during the first 300~Myr, when the SFR is higher, whereas Hdc never exceeds $\beta_{\rm out}\sim 5\times 10^{-3}$, and in many cases the outflow rate is unresolved.
In the bottom panels, we show the outflow velocity, computed as 
\begin{equation}
\langle v_{\rm out} \rangle = \frac{\sum_j m_j |v_{j,\rm z}|}{\sum_j m_j},
\end{equation}
where $m_j$ is the $j$-th particle mass and the sum is over the particles in the outflow ($v_z\cdot z >0$ and $|z-z_0|<\Delta z$, with $z_0$ the distance from the galactic plane). As a reference, we also show the escape velocity\footnote{The escape velocity is computed from the potential in the initial conditions, at the corresponding height above the disc plane and cylindrical radius $R=0$. We do not consider changes in the distribution during the simulation.} from the system, computed from the corresponding height $z$ above the disc to infinity, as a grey dot-dashed line. Also in this case, at 2~kpc Hmech results in slightly larger outflowing velocities relative to Hdc, within a factor of 2--3, because of the stronger effect of SN feedback. At larger distances, instead, the average outflow velocity is consistent between the runs, with Hdc showing a larger scatter because of the low gas mass involved, hence the under-sampling of the outflow properties. 
In conclusion, the mechanical feedback appears to be more effective in producing outflows than delayed-cooling, but at the same time it is not as effective in suppressing SF within the galaxy. This suggests that delayed-cooling, despite being empirical, probably mimics additional non-thermal processes which help suppressing SF in galaxies, whereas the mechanical feedback only accounts the `thermally driven' evolution of the bubble. A recent study by \citet{geen15} has shown that the preprocessing of the ISM in the presence of stellar radiation and young stellar winds can significantly alter the SN evolution, and can result in a larger terminal momentum of the bubble and a more effective suppression of SF. However, this is beyond the scope of this paper and will be investigated in the future.

\subsection{Resolution dependence of the scheme}
Now, we assess the resolution dependence of the mechanical feedback model, comparing five runs with different resolutions, i.e. VLmech,Lmech,Mmech,Hmech, and VHmech. 

\begin{figure}
\centering
\includegraphics[width=0.88\columnwidth]{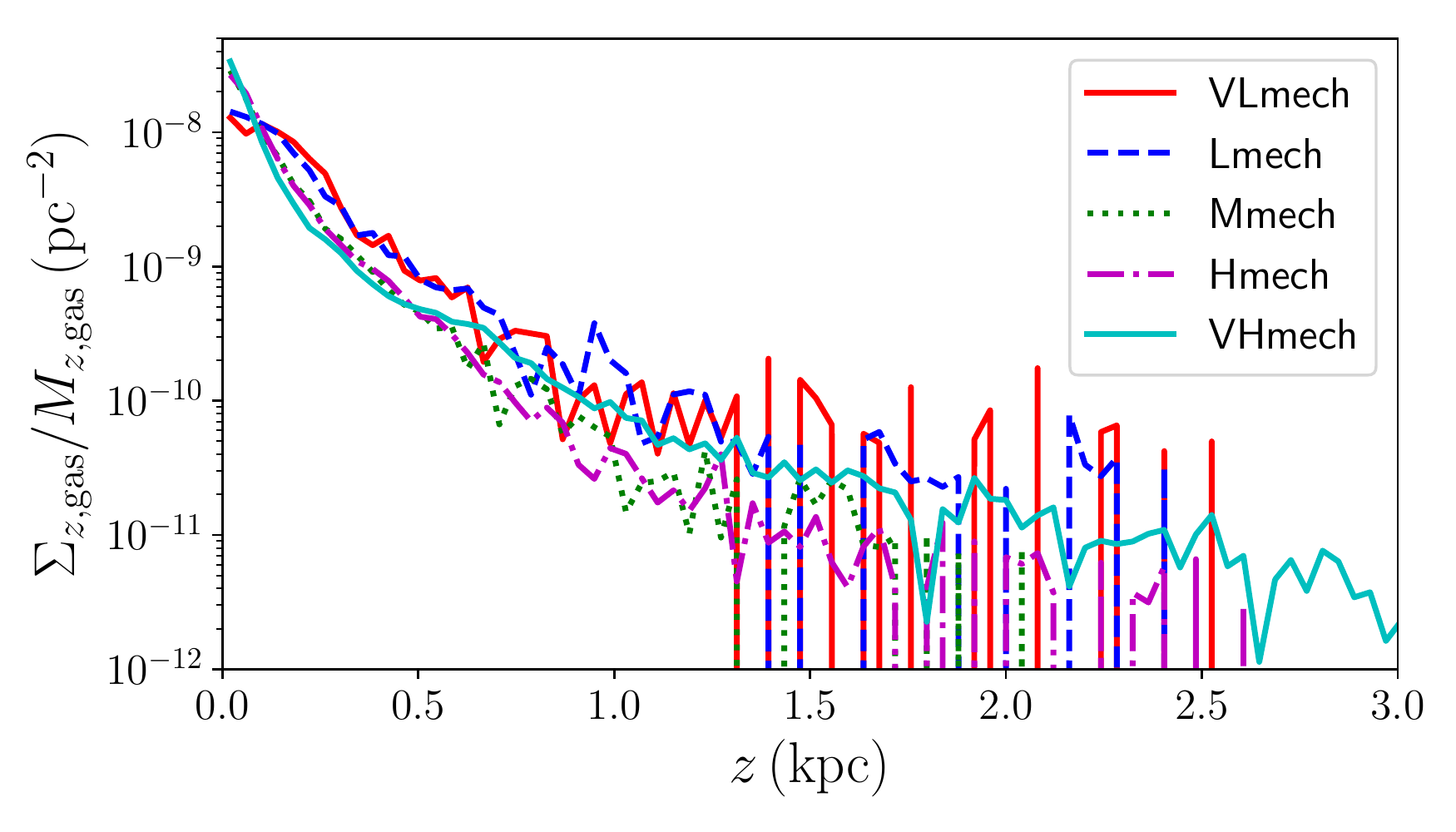}
\caption{Vertical surface density profile at $t=500$~Myr for five different resolutions. At low resolution, the disk is slightly thicker, because of the typically larger volume per particle and the subsequent stronger impact of SN feedback onto the gas farther away from the disc, where the densities are typically lower. As the resolution increases, the disc gets thinner, as observed in Mmech, Hmech, and VHmech, with a very good agreement in the profile within the central 500~pc among the higher resolution runs. At larger distances, where the density is very low, the differences in the particle sampling and the exact SF histories result in moderate discrepancies among the runs, with no clear trend with resolution.}
\label{fig:ressigma}
\end{figure}

\begin{figure}
\centering
\includegraphics[width=0.88\columnwidth]{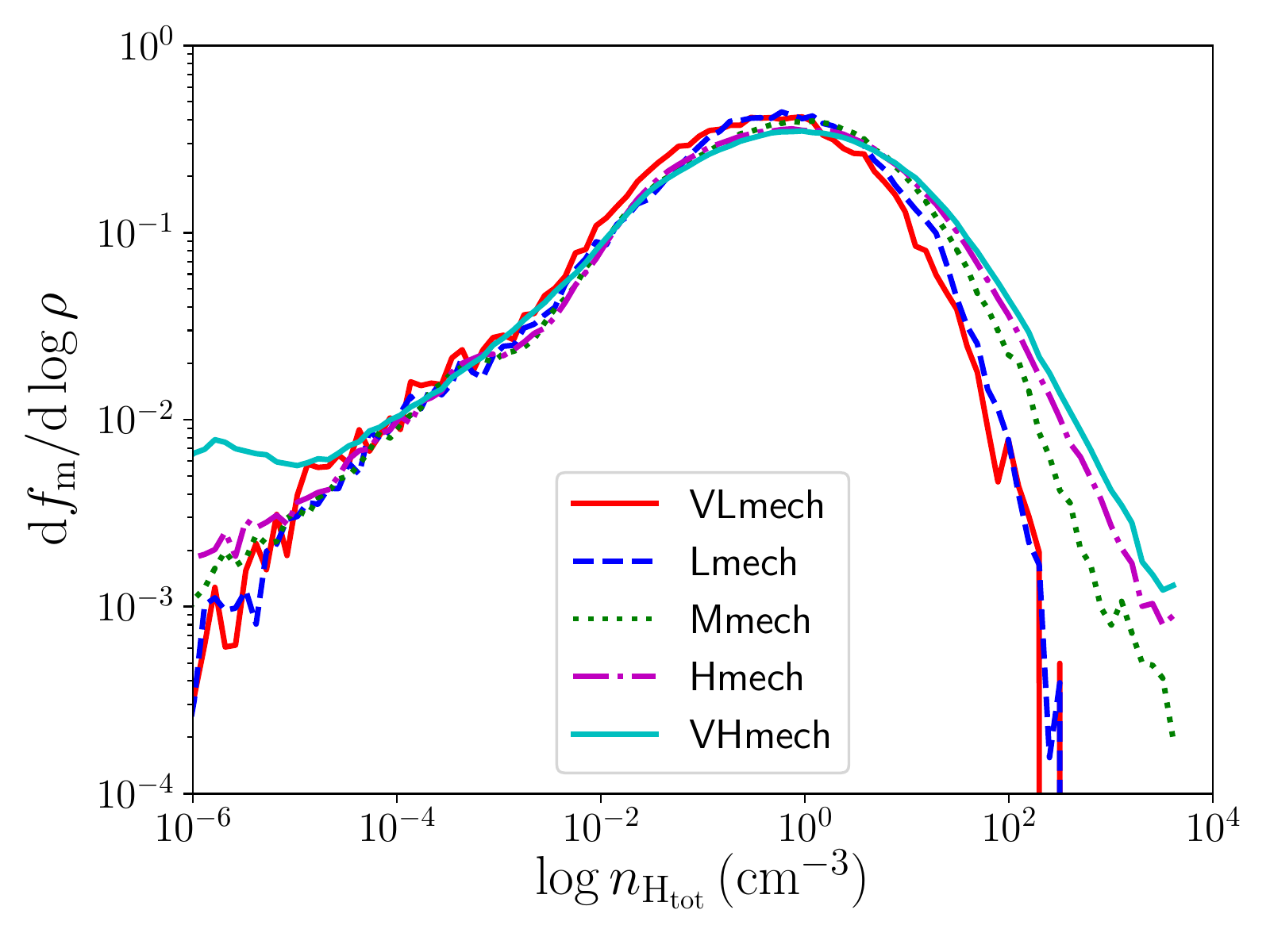}
\caption{Gas mass distribution for five different resolutions at $t=500$~Myr. The profiles are in very good agreement for $\nhtot\lesssim 1\rm\, cm^{-3}$, with the only exception being VHmech, where the extremely high resolution results in a slightly more massive tail at very low density (a factor of 3). At higher densities ($\nhtot\sim 10\rm\, cm^{-3}$), the better sampling of the velocity dispersion and the lower particle masses resulting from increasing resolution, allow us to better follow the collapse of the gas to higher densities before it is converted into stars.}
\label{fig:resphase}
\end{figure}

\begin{figure}
\centering
\includegraphics[width=0.9\columnwidth]{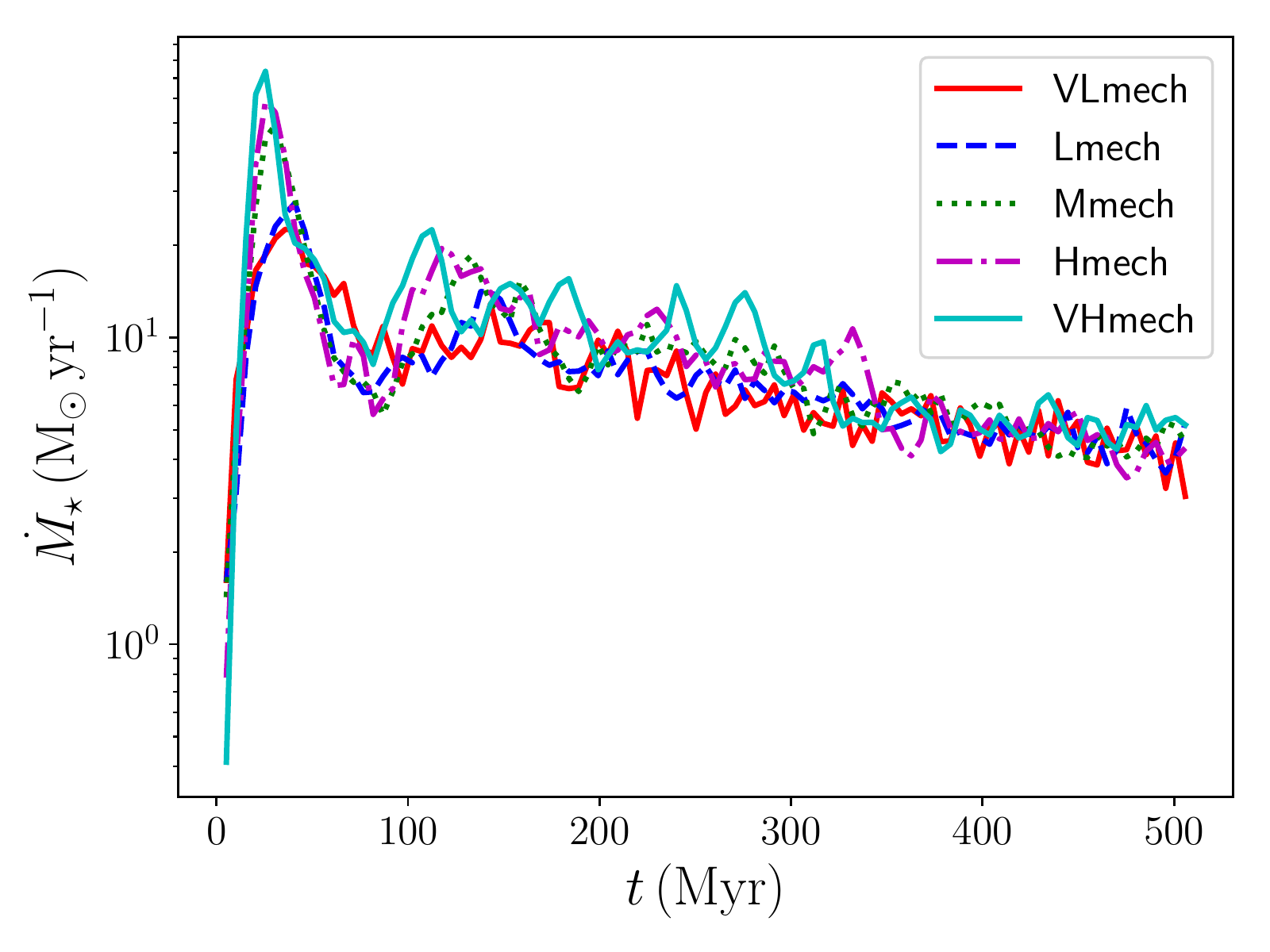}
\caption{SF rate for the five resolution cases. As the resolution increases, the initial peak gets higher, because of the instantaneous cooling that triggers the vertical collapse of the disc and the fragmentation on smaller and smaller scales. This is obvious when we compare Lmech with Mmech, whereas this effect gets suppressed when we increase further the resolution (see, i.e., Hmech and VHmech). Nevertheless, the subsequent SN events realign all the simulations, resulting in a consistent SFR rate evolution at later times.}
\label{fig:ressfr}
\end{figure}

\begin{figure*}
\includegraphics[width=0.82\textwidth]{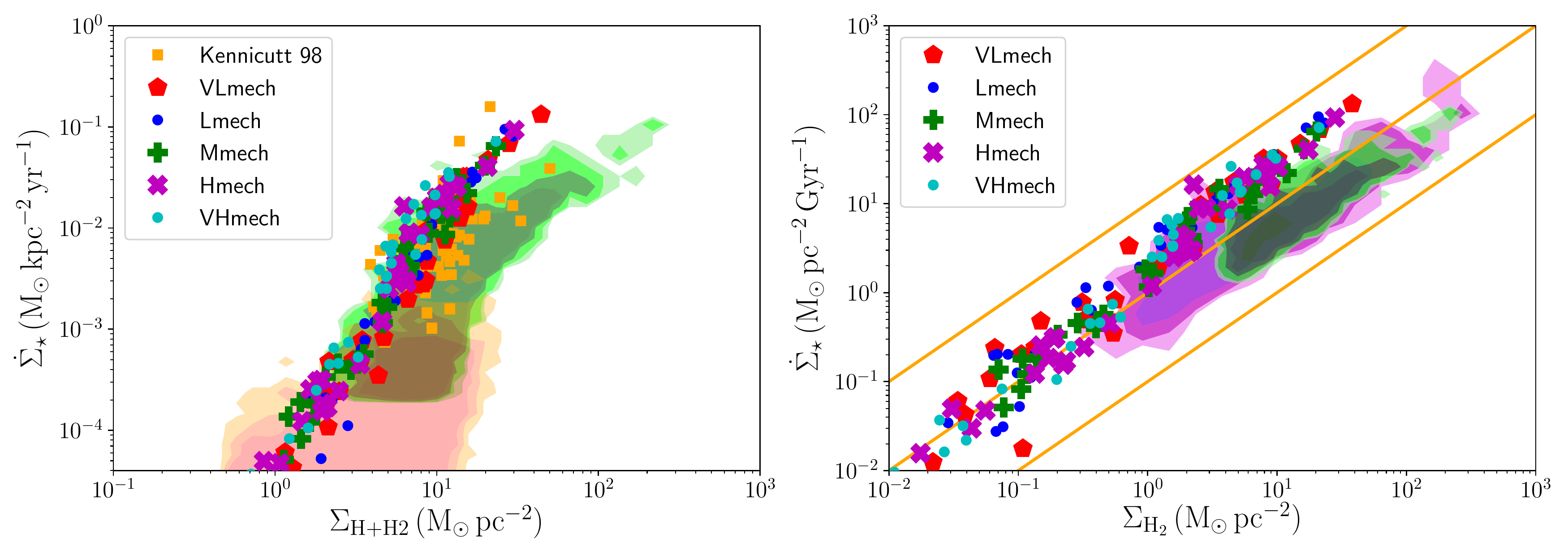}
\caption{KS relation for the five different resolutions at $t=500$~Myr, in both total gas (H+H$_2$, left-hand panel) and molecular gas (H$_2$, right-hand panel) only, as in Fig.~\ref{fig:dvsmks}. Despite the small differences in the evolution, all the simulations agree very well with each other in both panels. These results suggest that the KS relation is robust against resolution. Although not showed, the slope of the relation above $10~\msun~pc^{-2}$ is consistent with 1.4, while the normalisation is higher, as already discussed.}
\label{fig:resks}
\end{figure*}

\begin{figure*}
\includegraphics[width=0.75\textwidth]{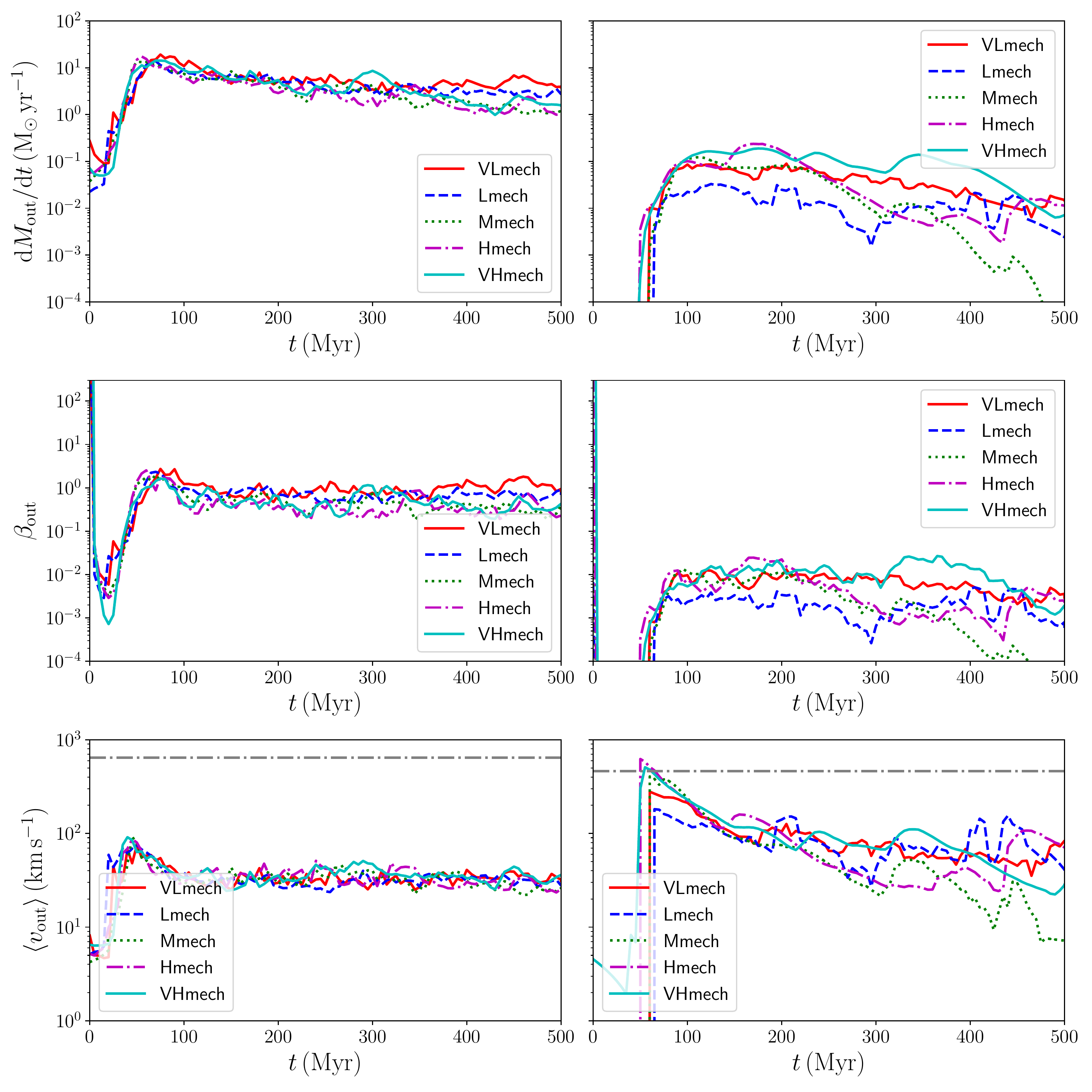}
\caption{Same as Fig.~\ref{fig:dvsmoutflow}, for the five runs at different resolutions. Closer to the disc plane, a coarser resolution favours the SN interaction with the gas further away and at slightly lower densities, resulting in mildly stronger outflows, whereas, as the resolution increases, the outflow rate and $\beta_{\rm out}$ tend to converge, in particular at late times. However, the typical velocity of the gas is not affected by the resolution. On the other hand, at larger distances, the moderate differences in the  evolution become more important, producing large stochastic variations in the outflow rates among the runs, but without a clear trend (for instance, VLmech and Hmech show outflow rates higher than Lmech or Mmech, whereas VHmech agrees well with Hmech in the first 100~Myr, then stays high, and finally realigns with Hmech in the last 50~Myr). In general, at later times, the net weakening of the outflows results in even larger differences.}
\label{fig:resoutflow}
\end{figure*}

In Fig.~\ref{fig:ressigma}, we report the vertical profile of the galaxy at different resolutions, following the same approach described in Fig.~\ref{fig:dvsmsigma}. At low resolution, the lower densities and larger volumes associated with each gas particle make the impact of SN explosions slightly stronger than at higher resolution, producing a thicker disc. In particular, if we define the height of the disc as the distance where $\Sigma_{\rm z,gas}$ is half the central peak, we get a value in the range $\sim 0.1-0.25$~kpc, with the minimum corresponding to the three high-resolution runs, and the maximum to the two low-resolution ones.

In Fig.~\ref{fig:resphase}, we compare the gas mass distribution as a function of density. In this case, the runs show a very similar behaviour, with an almost perfect overlap for $n_{\rm H_{tot}} \lesssim 1\rm\, cm^{-3}$, with the exception of a slightly more important tail at very-low density for VHmech, due to the extremely high resolution. At higher densities, instead, resolution effects become more important, with larger particle masses and the poorer sampling of the velocity dispersion resulting in a slightly more efficient SF already around $10\rm\, cm^{-3}$, hence preventing the gas from collapsing further. In addition, SNe become more effective when they explode in lower density environments, hence further reducing the gas in this density interval. As we increase the resolution, we start to better resolve higher densities.
Although not shown here, we find that the density--temperature diagrams are in very good agreement, except for the poorer sampling as we decrease the resolution, whereas the H$_2$ abundance is moderately lower, consistent with the slightly lower densities observed.

By comparing the SF history in the five runs in Fig.~\ref{fig:ressfr}, we can notice that the two low-resolution runs exhibit a mildly lower SFR in the first 250~Myr, consistent with the fact that i) the maximum density achieved is typically lower than in the high-resolution runs, thus reducing the initial burst of SF, and ii) SN events are more effective in sweeping away the gas when the density around the star is lower. Nevertheless, this difference almost vanishes as the simulation proceeds, resulting in a comparable SFR with small fluctuations. As we increase the resolution, these differences become less important, and the SFRs show a better agreement (see, for instance, Mmech and Hmech).

In Fig.~\ref{fig:resks}, we show the KS relation for the five runs in both total gas and H$_2$ only. Despite the large difference in resolution, the correlations in both panels are almost insensitive to the resolution. However, as already discussed, while the slope of the relations is in good agreement with the data, the normalisation of the KS relation in total gas is offset, because of the globally higher SFR in the simulations compared to the delayed-cooling model. These results suggest that the KS relation, especially the H$_2$-only counterpart, is robust against resolution.

Finally, we compare the outflow properties at different resolutions in Fig.~\ref{fig:resoutflow}. Closer to the disc plane, the outflows are a bit stronger for Lmech and VLmech (top--left and middle--left panels), because of the lower resolution and the SNe interacting with farther away neighbours. Nevertheless, all the runs agree with each other within a factor of a few difference, especially at earlier times, when the SFR is higher, hence the outflow rate. The average velocity is the same in all the runs. At larger distances, instead, we observe stronger differences (right-hand panels), but without a clear trend with resolution (for instance, in the last 100~Myr, VHmech is the highest, and Mmech the lowest). This suggests that these differences are most likely due to differences in the galaxy evolution (and the stellar distribution) that result in stochastic variations that are more evident at large distances where the outflow rate is generally lower. However, a convergence is far from being reached. 

\section{Discussion and conclusions}
\label{sec:conclusions}
We presented here a variation of the physically motivated SN feedback model presented in H18, that is able to deposit the right amount of energy and momentum into the ISM, according to the results of high-resolution, small-scale simulations. Most of the numerical details closely follow H18, but with some significant modifications, which include the maximum coupling radius, the effective face shared by particles, and the momentum/thermal energy deposition. After validating the model using isolated SN explosions in a uniform medium at different resolutions as done in H18, we applied the model to a Milky Way-like galaxy at $z\sim 0.1$ in isolation, to assess its ability to reproduce the observed correlation between SF and H$_2$. We compared this new model with the delayed-cooling SN feedback model employed in \citet{lupi18a}, which showed very good agreement with observations \citep[see also][]{capelo18}, despite the unphysical temperature of the gas kept hot by SNe, and we also tested the resolution dependence of the model.
We summarise here our findings:
\begin{itemize}
\item Terminal momentum in a uniform medium: the mechanical feedback model almost perfectly reproduces the terminal momentum of a single SN event, independent of resolution. However, to achieve this convergence, the proper swept-up fraction of the particle mass should be employed for the momentum deposition, otherwise the terminal momentum would be overestimated at high resolution. The delayed-cooling prescription, on the other hand, overestimates the terminal momentum by up to an order of magnitude, at low resolution, because of the cooling shut-off that draws the evolution closer to the fully energy-conserving case.
\item Model comparison -- SF in the galaxy: the delayed-cooling model more effectively suppresses SF, because of the artificial pressurisation of the gas, that completely inhibits SF for some time. When the shut-off time is short, the SF is not dispersed like in the mechanical feedback case, but it stays warm and not SF. In the opposite case, when the shut-off time is long enough, the SF region is destroyed, but the gas is kept hot, and it does not collapse efficiently when shocks occur with the surrounding medium. On the other hand, mechanical feedback more effectively sweeps the gas away, because of the kick imparted to the gas, but is not able to properly describe the hot, low-density cavity produced ( the gas remains cold and moves super-sonically, resulting in shocks and subsequent cooling that trigger new SF, yielding a slightly too high SF rate.
\item Model comparison -- KS relation: the H$_2$ KS relation and the H$_2$ column density fraction are consistent between the two models, with only mild differences, suggesting that these correlations are reasonably robust against different sub-grid models, as already stated in \citet{lupi18a} for the SF prescription. On the other hand, the total KS relation is higher in the mechanical feedback run, because of the overall higher SFR relative to the delayed-cooling case, although the slope in the high surface density regime is consistent with observations.
\item Model comparison -- galaxy outflows: close to the galaxy, the delayed-cooling model produces slightly stronger outflows ($\beta_{\rm out}$ is slightly larger than with mechanical feedback, but with a less than a factor of two difference), whereas the opposite occurs at larger distances. In all cases, the outflows never exceed the escape velocity from the halo, hence resulting in galactic fountains rather than proper outflows. This is consistent with previous results from \citet{rosdahl17}, where they find that only kinetic feedback is able to produce very powerful outflows. However, the idealised setup of our experiment does not allow to conclude whether mechanical feedback can explain the missing baryons in a cosmological context or not.
\item Resolution dependence: the mechanical feedback model is not strongly resolution-dependent. In particular, the outflows close the galactic plane (2~kpc) are similar among the runs, with only the low-resolution cases producing slightly stronger outflows because of the typically larger volumes of each cell and the SN--gas interaction that extends to larger distances from the disc plane. At larger distances, the variations increase up to one or two order of magnitudes among the different runs, but a clear trend cannot be observed, suggesting that the outflow rate evolution becomes very sensitive to the actual dynamical evolution of the galaxy and the location of the stellar particles, but not necessarily to the resolution itself. All the other properties, instead, are consistent among the five resolutions, with only larger fluctuations at lower resolutions because of the poorer sampling of the gas properties.
\end{itemize}

In the delayed-cooling model, one of the problems is the overshooting of the H$_2$ KS relation at high density. As already stated in \citet{lupi18a}, this could be related to (i) the fact that, at high densities, the SF prescription converts gas into stars very quickly, removing all the molecular hydrogen available for the converted gas particle, hence reducing its surface density, or (ii) to the numerical diffusion close to the resolution limit which slightly reduces the turbulent support of the gas. A possible solution to avoid (ii) is to include a sub-grid model for the turbulent cascade, as in \citet{semenov17}. However, from the comparison with the mechanical feedback, an alternative explanation could be that the high-density gas is kept warm by SNe, and H$_2$ is dissociated too efficiently. 

For the mechanical feedback model, instead, the higher SFRs observed can be the result of missing additional feedback processes like HII regions, young stellar winds, and cosmic rays, that could help provide additional energy/pressure able to more efficiently suppress SF. In the delayed-cooling runs, the empirical choice of shutting off radiative cooling could probably mimic (despite not in a completely physical fashion) these additional processes not present in the mechanical feedback model. An alternative explanation for the too high SFR in the mechanical feedback model is that the momentum injected is too low compared to reality \citep[see, e.g.][]{keller14,gentry17}, probably because of numerically enhanced mixing \citet{gentry18}. If this is the case, then the terminal momentum should be higher, the shocks in the swept-away gas should be stronger and the gas heating more effective. As a consequence, this could lead to larger mass loading factors and in a more effective suppression of SF, as shown by \citet{semenov18b} where their fiducial boost to the momentum is enough to perfectly match the KS relation (in both total and molecular gas).
However, a thorough investigation of these effect is beyond the scope of this study.

Concluding, although the delayed cooling has many advantages that make it suitable for galaxy-scale simulations, it is unable to properly reproduce all the galaxy properties observed, especially those dependent on the thermochemical state of the gas. On the other hand, the mechanical feedback, despite not being as effective as the delayed cooling in suppressing the SF, better reproduces the thermo-chemical state of the gas, the H$_2$ column densities observed, and is also able to drive more powerful outflows. Although further investigations are certainly necessary to fully assess the limitations of the mechanical feedback model, its physical motivation and ability to accurately reproduce the terminal momentum independent of resolution represent a step forward compared to the empirical delayed cooling mode. Thanks to the general effort of the community in improving the sub-grid modelling, mechanical feedback, coupled with other state-of-the-art prescriptions for processes like SF, chemistry, and enrichment will definitely help us to better understand how galaxies self-regulate during their evolution.
\section*{Acknowledgements}
We thank the anonymous referee for useful suggestions which helped improve the study.
We thank Marta Volonteri, Massimo Dotti, Pedro R. Capelo and Philip F. Hopkins for useful discussions and suggestions.
We thank Sijing Shen for having provided the metal cooling table described in \citet{shen10,shen13}.
AL acknowledges support from the European Research Council (Project No. 267117, `DARK'; Project no. 614199, `BLACK'). 
This work was granted access to the High Performance Computing resources of CINES under the allocations x2016046955, A0020406955, and A0040406955 by GENCI, and it has made use of the Horizon Cluster, funded by Institut d'Astrophysique de Paris, for the analysis of the simulation results. 

\bibliographystyle{mnras}
\bibliography{./Biblio}

\appendix
\section{Limiter for momentum coupling}
\label{app:limiter}

\begin{figure}
\centering
\includegraphics[width=\columnwidth]{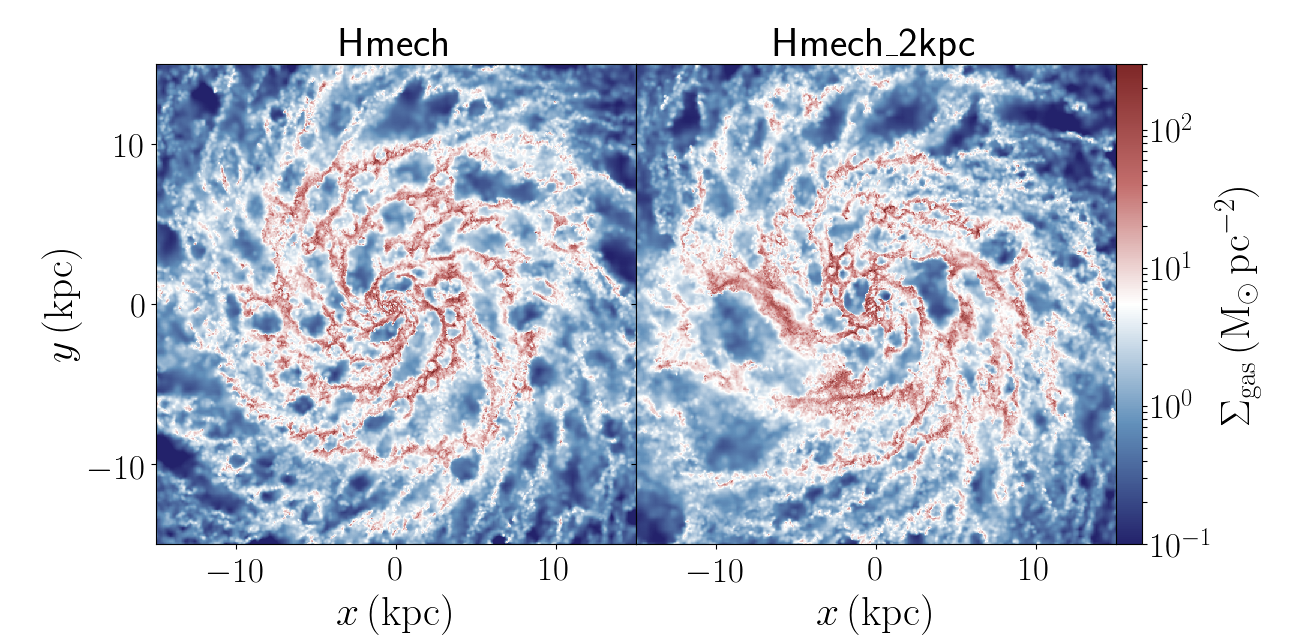}
\caption{Face-on view of the galaxy in Hmech (left-hand panels) and Hmech\_2kpc (right-hand panel) at the end of the simulation, as in Fig.~\ref{fig:dvsmmap}. In Hmech\_2kpc, the bubbles are typically larger, because of the larger coupling radius assumed. However, this is only a qualitative aspect, and it does not affect the quantitative evolution of the galaxy.}
\label{fig:2Kmap}
\end{figure}

\begin{figure}
\centering
\includegraphics[width=0.9\columnwidth]{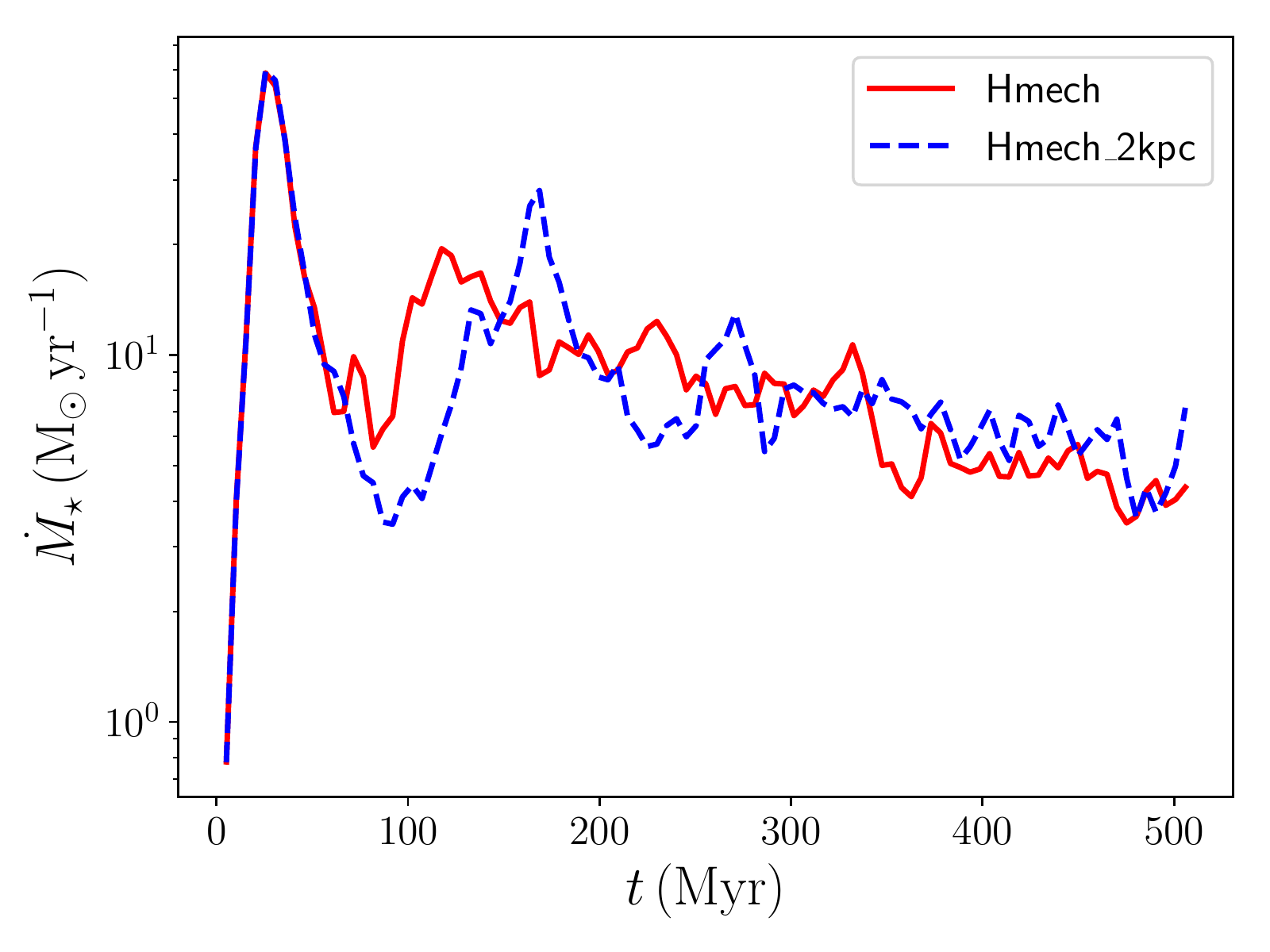}
\caption{SFR of the galaxy in Hmech (solid red) and Hmech\_2kpc (blue dashed). In Hmech\_2kpc, the initial burst is more effective in suppressing the SF, but it only shifts the second SF peak at later times, without significantly changing the final stellar mass (within a few per cent).}
\label{fig:2KSFR}
\end{figure}

\begin{figure*}
\centering
\includegraphics[width=0.79\textwidth]{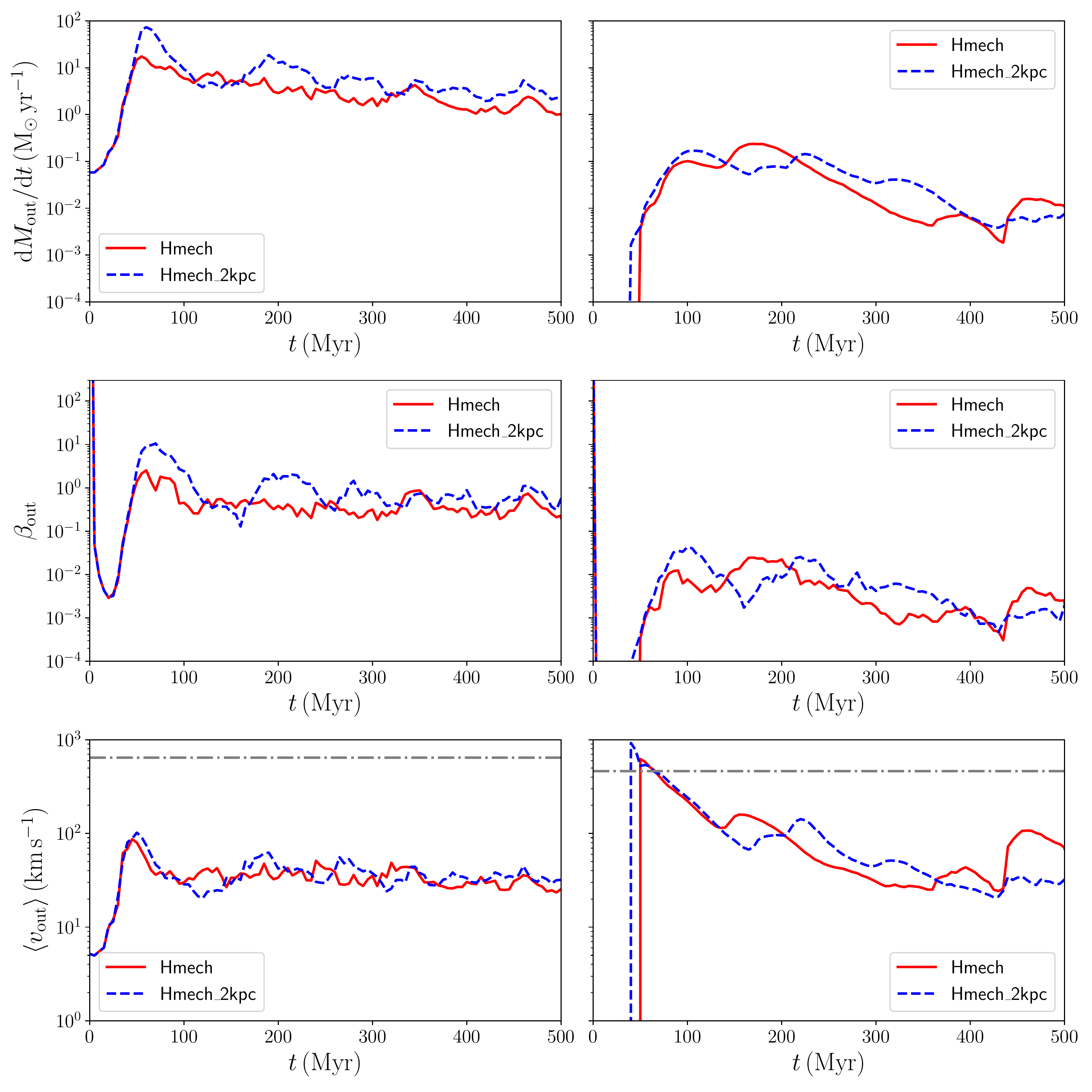}
\caption{$\beta_{\rm out}$ at 2~kpc and 20~kpc for Hmech (solid red) and Hmech\_2kpc (blue dashed). In Hmech\_2kpc, the larger coupling radius is able to push the gas more effectively at 2~kpc. However, at these distances from the disc, the exact choice of the maximum radius can play a role, although moderate. At larger distances, instead, this choice is less relevant, as suggested by the fluctuations between the two runs.}
\label{fig:2Kout}
\end{figure*}

Here, we discuss our choice of a distance/density limiter for the momentum coupling scheme, by presenting a variant of our mechanical feedback model, that we name `Hmech\_2kpc', where we couple the momentum obtained from Eq.~\eqref{eqn:ethpr} to all the particles within a distance of 2~kpc from the star (as done in H18), without any check on the density or on the distance of the effective face. As already stated in Section~\ref{sec:momcons}, for a very-low density medium, the SN bubble should merge while still in the energy conserving phase. In this case, the solution from \citet{martizzi15} is not correct any longer and can artificially enhance the feedback effect. To highlight how important the exact choice for the limiter is, we compare Hmech with Hmech\_2kpc. In Fig.~\ref{fig:2Kmap}, we show a face-on map of the gas in the two runs. Hmech\_2kpc, because of the larger coupling radius, can produce larger bubbles compared to Hmech. However, this does not affect significantly the other galaxy properties. For instance, the vertical profile of the disc is in perfect agreement between the two runs, although not reported. In Fig.~\ref{fig:2KSFR}, we show the SFR evolution for the two runs. We observe different fluctuations, because of the small differences in the coupling, in particular during the initial phases when the galaxy has not reached yet a steady-state evolution. At late times, instead, the two runs agree better. In Fig.~\ref{fig:2Kout}, instead, we compare $\beta_{\rm out}$ for the two runs. At 2~kpc (left-hand panel),  Hmech\_2kpc is slightly  above Hmech, because of the coupling extending up to larger distances. However, at 20~kpc, this difference becomes less important, with the two runs fluctuating significantly without a clear trend. The other quantities discussed in the main text are not reported here, for simplicity, because no differences have been observed between the two runs. We can therefore conclude that the exact choice of the limiter does not play a significant role in the galaxy evolution, as long as the value is large enough (for instance, a multiple of the expected merging radius at very low density) to avoid spurious suppression of the feedback.

\label{lastpage}
\end{document}